\documentclass[aps,prc,floats,floatfix,twocolumn,showpacs,superscriptaddress,nofootinbib]{revtex4}

\usepackage{epsfig}
\newcommand{\nuc}[2]{$^{#1}${#2}}
\newcommand{\etal}{\emph{et al.}}
\newcommand{\nn}{\nonumber}

\begin{document}

\title{Global study of quadrupole correlation effects}

\author{M. Bender}
\affiliation{Department of Physics and Institute for Nuclear Theory,
             Box 351560, University of Washington, Seattle, WA 98195}
\affiliation{Physics Division, Argonne National Laboratory,
             9700 S. Cass Avenue, Argonne, IL 60439}
\affiliation{National Superconducting Cyclotron Laboratory,
             Michigan State University, East Lansing, MI 48824}

\author{G. F. Bertsch}
\affiliation{Department of Physics and Institute for Nuclear Theory,
             Box 351560, University of Washington, Seattle, WA 98195}

\author{P.-H. Heenen}
\affiliation{Service de Physique Nucl{\'e}aire Th{\'e}orique,
             Universit{\'e} Libre de Bruxelles,
             CP 229, B-1050 Brussels, Belgium}

\date{August 26 2005}


\begin{abstract}
We discuss the systematics of ground-state quadrupole correlations
of binding energies and mean-square charge radii for all even-even
nuclei, from \nuc{16}{O} up to the superheavies, for which data 
are available.
To that aim we calculate their correlated \mbox{$J=0$} ground state
by means of the angular-momentum and particle-number projected generator
coordinate method, using the axial mass quadrupole moment as the
generator coordinate and self-consistent mean-field states only
restricted by axial, parity, and time-reversal symmetries.
The calculation is performed within the framework of a
non-relativistic self-consistent mean-field model using the same
non-relativistic Skyrme interaction SLy4 and a density-dependent
pairing force to generate the mean-field configurations and mix
them.
The main conclusions of our study are:
(i) The quadrupole correlation energy varies
between a few 100 keV and about 5.5 MeV. It is affected by shell
closures, but varies only slightly with mass and asymmetry.
(ii) Projection on angular momentum \mbox{$J=0$} provides the major
part of the energy gain of up to about 4 MeV; all nuclei in the study
including doubly magic gain energy by deformation.
(iii) the mixing of
projected states with different intrinsic axial deformation adds
a few 100 keV up to 1.5 MeV to the correlation energy.
(iv) Typically nuclei below mass \mbox{$A \leq 60$} have a larger
correlation energy than static deformation energy while the heavier
deformed nuclei have larger static deformation energy than correlation
energy.
(v) Inclusion of the quadrupole correlation energy improves the
description of mass systematics, particularly around shell closures
and for differential quantities, namely two-nucleon separation energies
and two-nucleon gaps. The correlation energy provides an explanation of
``mutually enhanced magicity".
(vi) The correlation energy tends to decrease the shell effect on binding 
energies around magic numbers, but the magnitude of the suppression is 
not large enough to explain the relative overbinding at \mbox{$N=82$} 
and \mbox{$N=126$} neutron shell closures in mean-field models.
(vii) Charge radii are also found to be sensitive to the quadrupole
correlations. Static quadrupole deformations lead to a significant
improvement of the overall systematics of charge radii. The dynamical
correlations improve the local systematics of radii, in particular
around shell closures. Although the dynamical
correlations might reduce the charge radii for specific nuclei, they
lead to an overall increase of radii when included, in particular in
light nuclei.
\end{abstract}
\maketitle
%
%
\section{Introduction}
\label{sect:intro}
The last few years witnessed a tremendous progress toward the
construction of microscopic models for nuclear masses. Most
promising for such an endeavor is self-consistent mean-field
theory (SCMF), also called density functional theory
(DFT). See Ref.\ \cite{RMP} for a recent review. Since
the work of Tondeur \etal\ \cite{Ton00a}, fits of self-consistent
mean-field models to all experimentally known masses have become
available. However, the fits are only competitive with the
\emph {de facto} standard of mass formulae, the finite-range liquid drop
model (FRDM) \cite{Mol95a}, when phenomenological corrections are
added for certain correlation effects.  Namely, a rotational
energy correction is added when the mean field state breaks
spherical symmetry, and a Wigner term is added to account for the
stronger neutron-proton correlation in nuclei having nearly equal numbers of
neutrons and protons \cite{Gor01a,Sam02a}. The quality of
the fits, either FRDM or mean-field + corrections, is in the range
of 0.6-0.7 MeV rms.

Some time ago, Bohigas and Leboeuf \cite{Boh02a} opened a
discussion on limits of accuracy of theories of the nuclear masses
by arguing that chaotic contributions to the nuclear wave function
will ultimately limit the accuracy of any mean-field approach.
Their rough estimate for this limit is a rms deviation of
$\sigma_{\text{rms}} \approx 500$ keV (see Eq.\
(\ref{eq:sigmarms}) for its definition), which is only slightly
below the values achieved in the FRDM and the mean-field +
corrections. On the other hand, applying noise analysis to
available mass theories of various types, Barea \etal\ argue in
Ref.\ \cite{Bar05a} that the upper limit of mass predictability
should be well below 100 keV. They find that global mass
theories such as the FRDM or the HFB show correlated
errors. Only the much better performing local mass
models, as for example the Garvey-Kelson (GK) mass formulas
\cite{Gar66a,Gar69a,Com88a} with a $\sigma_{\text{rms}}$ as low
as 86 keV, have residual errors which are consistent with white
noise. Another interesting recent analysis was made by Molinari
and Weidenm{\"u}ller \cite{Mol04a}, who computed the effect on the
ground state energy of the fluctuations associated with the
coupling to states at tens of MeV of excitation.  The estimate in
their Fig.\ 1 is about 100 keV, again much smaller than the
accuracy achieved by the global mass fits. Summarizing the current
status of the discussion about the limits of mass theories, it is
likely that the current limit of about 600 keV is ``not a physical
phenomenon, but rather a characteristic arising from the
mean-field approximation'', to quote the conclusion of Ref.\
\cite{Vel05a}.

Even the limit of 600 keV may be optimistic.  Indeed,
stripping away all phenomenological corrections, it was found in
Ref.\ \cite{Ber05a} that SCMF theories based on Skyrme parameterizations
only achieve a factor of two improvement over the liquid drop
model, yielding rms residuals in the range 1.5-1.7 MeV. The need for
explicit treatment of correlations should not be surprising.
A remarkable fact about nuclear structure physics is that the mean
field approach works as well as it does, given the strong short-range
character of the nucleon-nucleon interaction.  In this respect,
the nuclear problem is much more difficult than the problem of
structure and binding of electronic systems.  There, the
interaction is smooth and the self-consistent field is a very good
starting point.  What saves the theory for nuclear systems is the
fact that the correlations induced by the interaction largely have
a short range themselves. This makes it plausible that they can
be subsumed in an effective
interaction.  But not all correlation effects can be included by a
renormalization of a short-range interaction.  The electronic
problem provides a nice example of this.  In the usual Kohn-Sham
theory, the energy functional is local or nearly local except for
the kinetic energy and electrostatic interactions. This is quite
inadequate to describe the long-range van der Waals interaction,
which is absent in the mean-field wave function and thus a pure
correlation effect.

Another way that long-range correlations come about is when a
symmetry is broken in the mean field and in the corresponding
density matrix.  In the nuclear many-body problem, translational
symmetry is always broken and rotational symmetry may or may not
be broken depending on the nucleus.  The true ground state of
course has the symmetry restored, and the additions to the wave
function that bring about the symmetry restoration are correlation
effects that are necessarily long-range when the symmetry is a
global one. From the point of view of making a theory of the
masses, there are two important questions.  The first is how large
are the correlation energies associated with broken symmetries.
Unless their size is greater than the target accuracy of theory,
they can be ignored.  In fact, both the center-of-mass energy and
the rotational energy are large compared to the 600 keV
present-day standard.  The second question is how much the energy
fluctuates from nucleus to nucleus. If the correlation energy
varies very smoothly, it could remain unnoticed in a theory based
on a fitted energy functional.  This seems to be the case for the
center-of-mass energy. While its size can be of the order of 10
MeV for the lighter nuclei, its fluctuations are much less
important than those of the quadrupolar degrees of freedom (see
\cite{Ben00a}).

The situation is more precarious for the rotational energy.  It
fluctuates considerably from nucleus to nucleus, vanishing for
spherical nuclei and having a magnitude of the order of several
MeV for deformed nuclei.  This provides a motivation to calculate
this correlation energy explicitly rather than keeping it buried
as part of the mean-field energy functional. However, it is
dubious to treat it as a discrete quantity, present
in some nuclei but not in others. The shape can fluctuate and the
binary classification of spherical or deformed nuclei should be
replaced by a continuum starting from rigid spherical, through
soft transitional to statically deformed nuclei. Thus, one is led
to seek a theory of the correlation energy that would include the
energy associated with zero-point shape fluctuations as well as
static deformations.

There are two leading candidates for a systematic and practical theory
of long-range correlations effects taking mean field theory as the
starting point. These are the RPA, generalized to QRPA in the presence of
pairing, and the generator coordinate method (GCM). The
RPA has an impeccable pedigree in quantum many-body theory, solving a
long-range divergence problem in the calculation of the correlation
energy for Coulomb interactions. However, balancing this are
a number of drawbacks which we list:

\begin{itemize}
\item  RPA does not converge well when the interactions are short
ranged \cite{Ste02}.  This becomes obvious when one notes that second-order
perturbation theory for a contact interaction diverges, and that the
usual formula for the RPA correlation energy incorporates
the second-order perturbation.
\item
As a small-amplitude approximation, QRPA cannot be expected to give
a good description of the correlated ground state in soft transitional
nuclei and nuclei with coexisting minima, where the ground-state is
spread over a wide range of deformations.
\end{itemize}

None of these problems of RPA are necessarily insurmountable.
Concerning the convergence, one might explicitly exclude the
second-order perturbation term to eliminate the divergence.
Alternatively, one could regularize the interaction in some way.
Along these lines, the authors of Ref.\ \cite{Bar04} calculated RPA
correlation energies in the Sn isotope chain using nuclear field
theory, a theory that replaces the microscopic particle-hole
interaction with a surface-peaked multipole interaction. 
The criticism of RPA that it is limited to small amplitudes 
is not entirely justified in practice: it can treat the
correlation energy associated with symmetry restorations which are
large amplitude effects \cite{Rin80a}, although the quality of this
approximate symmetry restoration is not always very good \cite{Joh02a}.

The other leading contender for a theory of correlation effects is
the GCM, which we favor and apply in this work.  The essential
idea is that one considers a manifold of mean-field states in an
external field, different strengths of the external field
generating different states. The important point is that the space
is essentially determined by the functional form of the external
field.  Once the field or the set of fields is specified, the
theory is completely systematic and applicable to all nuclei for
which the mean field is a reasonable starting point.  The GCM is a
``horizontal" extended theory in that the added parts to the wave
function are low energy configurations, because they were obtained
by the mean-field minimization procedure. This contrasts to the
RPA approach, which invokes a "vertical" extension of the wave
function to arbitrary high energy, but to only states that can be
accessed by a one-body operator. Since we mentioned a number of 
drawbacks of (Q)RPA, we also make a similar list for the GCM. Namely:

\begin{itemize}
\item
Convergence can be an issue on a numerical level. The theory is
usually couched in terms of a continuum of mean-field states, but
in practice computations are carried out with finite sets of states.
If there are too many states in the basis of non-orthogonal states,
they will be redundant and the matrix techniques to find the lowest
energy state become unstable.
\item
For numerical reasons, we are limited at present to a single
external field. We take it to be the isoscalar axial quadrupole
field,
\begin{equation}
\label{eq:Q_2}
Q_2
= 2 z^2 -x^2 -y^2
\end{equation}
leaving out higher multipoles and non-axial quadrupolar
deformations.
\item
Already for \mbox{$J=2$} excitations, it may be insufficient to
take only the single generating field from Eqn.\ (\ref{eq:Q_2}).
This suspicion is raised by systematic overestimations of
quadrupole excitations in rigid spherical nuclei \cite{Rod02a,Ben03a}.
\end{itemize}

In the present paper we aim at a systematic study of the quadrupole
correlation energy for all even-even nuclei where the mass is known.
The quadrupole correlation can be expected to be the dominant
correlation mode for all but doubly magic nuclei, so this is at
least a reasonable first step to a complete theory.
Starting from a self-consistent mean-field
model based on Skyrme's interaction, we restore particle-number
and rotational symmetry and perform a configuration mixing of
states with different quadrupole moment. Our approach does not
aim at a nuclear mass formula, but has the long-range goal of
a universal nuclear model that
allows for the simultaneous consistent and systematic description of
many observables, including excited states, for all nuclei.
Such a strategy has obvious advantages for the most prominent
application of nuclear mass formula, the description of
nucleosynthesis in astrophysics. For example, the dynamics of the
$r$ process of explosive nucleosynthesis is determined by many
nuclear properties \cite{Lan01a,Arn03a}, with masses being
just the simplest one.

The paper is organized as follows: In section \ref{sect:calc}, we
present the equations to be solved and discuss some
technical aspects. In Section \ref{sect:examples}, 
the physics of quadrupole correlations is discussed on a few typical
examples.
A systematic
calculation for the correlation energies of 605 even-even nuclei is
presented in section \ref{sect:correlations:overview}
while in Sect.\ \ref{sec:mass:table:syst}, we analyze the effect of
correlation energies on mass residuals. In section \ref{sect:global}
we discuss the role of quadrupole correlation energies from the
point of view of mass models. In Sect.\ \ref{sect:radii}, we examine
the role of quadrupole correlations for charge radii. A summary and our
conclusions are presented in Sect.\ \ref{sect:summary}.
Some of the key results for quadrupole correlation energies have
been presented earlier in a letter \cite{Ben05a}.

%
%
\section{Calculational Procedure}
\label{sect:calc}

%
%
\subsection{Mean Field}
\label{subsect:calc:mf}

We start with a set of self-consistent solutions of the HF+BCS
generated with the code {\tt ev8} \cite{Bon85a,Bon05a}. The
single-particle wave functions are discretized on a
three-dimensional Lagrange mesh \cite{Bay86a} corresponding to a
cubic box. The only restriction of the wave function is that
the Slater determinant of the orbits is invariant with respect to
parity, time reversal, and axial rotations. As in earlier studies,
we use a fixed mesh spacing of 0.8 fm. The length of the box side
ranges from 25.6 to 28.8 fm with the nucleus at the center.  To
avoid the breakdown of pairing correlations for small level
densities around the Fermi surface, we perform an approximate
projection-before-variation on particle number within the
Lipkin-Nogami (LN) scheme as outlined in \cite{Gal94a}. States
with different mass quadrupole moments are generated by adding a
constraint to the mean-field equations to force the quadrupole
moment, Eqn.\ (\ref{eq:Q_2}), to have some value
\begin{equation}
q
= \langle Q_2 \rangle
.
\end{equation}
Higher even axial multipole moments are automatically optimized for a
given mass quadrupole moment. For numerical stability of the constrained
mean-field equations in light nuclei, the constraint is damped at
large distances from the nuclear surface with the method proposed
by Rutz \etal\ in Ref.\ \cite{Rut95}.

The Skyrme interaction SLy4 \cite{Cha98} is used for the 
energy-density functional in the particle-hole channel. Pairing effects 
are treated in the BCS approximation using a density-dependent zero-range
force, truncated above and below the Fermi surface as described in
Ref.\ \cite{Rig99}. As in earlier studies, the pairing strength is
taken to be $-1000$ MeV fm$^3$ for both protons and neutrons.

While the wave functions are constructed using the code {\tt ev8},
all energies and matrix elements are calculated with another code,
{\tt promesse} \cite{promesse}, which uses a more accurate
algorithm for the kinetic energy. For SCMF energies, this code has
an accuracy for a mesh size of 0.8~fm given roughly by 0.007 $A$
MeV, where $A$ is the mass number. This error varies quite
smoothly with $A$. For even better accuracy, the mesh spacing
should be decreased. Highly accurate SCMF calculations are also
achieved using a deformed harmonic oscillator basis; see Ref.\
\cite{st05b} for code details.

%
%
\subsection{Beyond the Mean Field}

The application of the generator coordinate method that we do here
goes beyond mean field in three respects: projections on good
particle numbers, projection on angular momentum \mbox{$J=0$}, and
mixing deformations. Projection is a special case of the GCM,
where the collective path and the weight functions are determined
by the restored symmetry. Angular-momentum projection  mixes
states with all the possible orientations of the quadrupole tensor
and therefore generates part of the quadrupole correlations. For
this reason, to introduce consistently quadrupole correlations,
the mixing of states with respect to the quadrupole moment by the
GCM should be performed together with an angular-momentum
projection.

Eigenstates of the particle-number operator $\hat{N}$ with an even
eigenvalue $N_0$ are obtained by applying the particle-number
projection operator
\begin{equation}
\hat{P}_{N}(N_0)
= \frac{1}{\pi}
  \int_{0}^{\pi} \! d \phi_N \;
  e^{i \phi_N (\hat{N}-N_0)}
,
\end{equation}
separately for protons and neutrons. All the results presented in this 
paper include particle-number projection and we drop the indices $N_0$ 
and $Z_0$ to simplify the notations. To avoid the use of the 
Lipkin-Nogami correction of mean-field energies which is known to be often
inaccurate, the mean-field energies corresponding to the Skyrme and pairing
interactions have been recalculated by projecting the SCMF states on
particle number. Thus, in the following, what we call SCMF energy is in 
fact the energy corresponding to a particle-number projected SCMF state.
This projection is always performed on mean-field states which have
$N_0$ and $Z_0$ as average particle numbers.

Formally, eigenstates $| J M q \rangle$ of the  angular momentum operators
$\hat{J}^2$ and $\hat{J}_z$ with eigenvalues \mbox{$J (J+1)$} and $M$
are obtained by applying the operator
\begin{equation}
\label{eq:PJ}
\hat{P}^J_{MK}
= \frac{2J+1}{16 \pi^2}
  \! \int_{0}^{4\pi} \! \! d\alpha
  \! \int_0^\pi \! \! d\theta \; \sin(\theta)
  \!  \int_0^{2 \pi} \! \! d\gamma \;
  \mathcal{D}^{*J}_{MK} \,
  \hat{R} 
,
\end{equation}
on the states $| q \rangle$. The rotation operator
$\hat{R}$ and the Wigner function $\mathcal{D}^{J}_{MK}$ both depend
on the Euler angles $\alpha,\theta,\gamma$.  In practice, we shall
simplify the 3-dimensional integral over Euler angles to a one-dimensional
integral (see Eqns.\ (\ref{eq:norm:kernel:exact}-\ref{eq:ham:kernel:exact})
below).

The second step in treating quadrupole correlations is to mix
configurations of different deformations.  The mixed projected
many-body state is set up as a coherent superposition of
normalized projected mean-field states $| J M q \rangle$ with
different intrinsic deformations $q$
\begin{equation}
\label{eq:p:gcm:state}
| J M k \rangle
= \sum_{q} f_{J k} (q) \, | J M q \rangle
.
\end{equation}
The weight function $f_{J,k} (q)$ is determined to minimize the
energy expression
\begin{equation}
\label{eq:GCM:E}
E_k
=\frac{\langle J M k | \hat{H} | J M k \rangle}
      {\langle J M k | J M k \rangle}
.
\end{equation}
where we have omitted the angular momentum indices on $E_k$ since
we will only be interested in the following to $J=0$ states.
The solution is given by a matrix eigenvalue equation that corresponds
to the discretized Hill-Wheeler-Griffin (HWG) equation \cite{Hil53a,Gri57a}
\begin{equation}
\label{eq:HWG}
\sum_{q'}
\big[ H_J (q,q') - E_k \, I_J (q,q') \big] \; f_{J,k} (q')
= 0
.
\end{equation}
For each $J$-value, the HWG equation gives a full spectrum of
correlated states corresponding to the collective variable $q$.
This spectrum can be used to study collective excitations, see,
e.g., \cite{Egi04a,Ben05b} and references given therein. Here we are
only interested in the \mbox{$J=0$} ground state.
Note that the exact number projection avoids some problems that may
arise in calculating HWG matrix elements with BCS wave functions \cite{Egi04a}.

The angular momentum projection simplifies to a one-dimensional
integral when the mean-field states are axial and time-reversal
invariant. We take the $z$ axis as a symmetry axis  and the
mean-field states as eigenstates of the $z$ projection of the
angular momentum in the intrinsic frame, with eigenvalue
\mbox{$K=0$}. The study of only even-even nuclei permits us to use a
reduced interval for the angular integration. The
angular-momentum projected norm and Hamiltonian kernels entering
Eqn.\ (\ref{eq:HWG}) are then given by
\begin{eqnarray}
\label{eq:norm:kernel:exact}
I_J (q,q') & = & \langle J M q | J M q' \rangle
      \nn \\
& = & \frac{1}{\mathcal{N}_J (q) \mathcal{N}_J (q')} \!
      \int_0^{\pi/2} \! \! d \theta \: \sin (\theta) \, d^J_{00} (\theta) \,
            \langle q | \hat{R} (\theta) | q' \rangle
      \nn \\
      \\
\label{eq:ham:kernel:exact}
H_J (q,q') & = & \langle J M q | \hat{H} | J M q' \rangle
      \nn \\
& = & \frac{1}{\mathcal{N}_J (q) \mathcal{N}_J (q')} \!
      \int_0^{\pi/2} \! \! d \theta \: \sin (\theta) \, d^J_{00} (\theta) \,
             \langle q | \hat{R} (\theta) \hat{H} | q' \rangle
.
      \nn \\
\end{eqnarray}
with
\begin{equation}
\mathcal{N}_J (q) 
= \sqrt{\int_0^{\pi/2} \! d \theta \: \sin (\theta) \, d^J_{00} (\theta) \,
            \langle q | \hat{R} (\theta) | q' \rangle} .
\end{equation}
The description of odd and odd-odd nuclei would
require to break time reversal and axial symmetries \cite{Dug02a}, 
increasing the complexity of the calculation by several orders of 
magnitude.

Note that the weight functions $f_{J,k}(q)$ in Eqn.\
(\ref{eq:p:gcm:state}) are not orthogonal. A set of orthonormal
collective wave functions \mbox{$g_{J k} (q)$} in the basis of 
the intrinsic states is obtained by a transformation involving 
the square root of the norm kernel \cite{Rin80a,RMP}.

The expressions given above are written for a many-body
Hamiltonian, $\hat{H}$. In practice, however, we use an energy
density functional for the effective interaction, replacing all
densities in the functional are by their corresponding transition
densities.

In terms of the computational algorithms, an important technical
challenge of a configuration-mixing calculation is the computation
of the non-diagonal matrix elements between mean-field states.
These are evaluated with the help of  a generalized Wick theorem
\cite{Bal69a}. The single-particle states are discretized  on a
3-dimensional mesh in coordinate space using a Lagrange mesh
technique \cite{Bay86a}. Thanks to the imaginary time step method
\cite{Dav80a}, only a small fraction of the single-particle states
which could be constructed on the mesh need to be computed. As a result,
the two sets of single-particle states corresponding to two
mean-field solutions are not equivalent, which has to be carefully
taken into account \cite{Bon90a,Val00a}. The overlaps are
calculated with the Onishi formula. It contains a square root
evaluation, which has a sign ambiguity that requires some
additional care.

Another technical problem that appears at the level of solving
the HWG equation (\ref{eq:HWG}) is the possible over-completeness of the
basis states $|q\rangle$. This can lead to problems of numerical 
stability, see Sect.\ \ref{sect:mixing} below.

%
%
\subsection{Definition of correlation energies}
The energy of an angular-momentum projected mean-field state
of deformation $q$ is given by the diagonal matrix elements
of the Hamiltonian kernel (\ref{eq:ham:kernel:exact})
\begin{equation}
E_{0}(q) 
= H_0(q,q)
.
\end{equation}
We denote the energy of a SCMF state $|q \rangle$ as $E(q)$. The 
energy gained by the projection of a state $|q \rangle$ is its 
\emph{rotational energy}
\begin{equation}
E_{\text{rot}}(q) 
= E(q) - E_0(q) 
.
\end{equation}
This energy can be computed in approximate ways without getting
into the details of a full projection \cite{Egi04a}.

Starting from the SCMF energy landscape $E(q)$, several correlation 
energies can be defined. The static \emph{deformation energy} is the energy
difference between a mean-field configuration $q$ and the spherical
one
\begin{equation}
E_{\text{def}}(q) 
= E(Q_2 = 0) - E(q) 
.
\end{equation}
The minimum of $E(q)$ is the SCMF energy, $E_{\text{mf}}$, and 
corresponds to a deformation $q_{\text{mf}}$.

After angular momentum projection, the minimum of $E_0(q)$ may 
correspond to a different configuration $q$ that we label $q_0$.
It is more useful to define correlation energies which can be
simply added to the SCMF energy binding energy. For angular-momentum 
projection, we introduce the \emph{rotational energy correction}:
\begin{eqnarray}
\label{eq:E:J=0}
E_{J=0} 
& = & E(q_{\text{mf}}) - E_0(q_0)
      \nn \\
& = & [E(q_{\text{mf}}) -E(q_0)] + E_{\text{rot}}(q_0) 
,
\end{eqnarray}
in which the first term represents a loss of energy due to
mean-field deformation and the second (larger) term the gain due
to angular momentum projection. 

The correlation energy gained by configuration mixing is then 
defined with respect to $E_{J=0}$:
\begin{equation}
\label{eq:E:GCM}
E_{\text{GCM}} 
= E_0 (q_0) - E_{k=0} 
.
\end{equation}
Both $E_{J=0}$ and $E_{\text{GCM}}$ are non-negative since they are
determined by a variational calculation. As a consequence, the
nucleus is always more bound by correlations. $E_{J=0}$ is also 
non-negative in other more approximate treatments of projection. 
However, when treating configuration mixing through a Bohr 
Hamiltonian or a collective Schr{\"o}dinger equation, the potential 
energy surface has a different meaning. It contains a (local) 
vibrational term (strangely named "zero point energy correction") 
which has the same sign as $E_{\text{GCM}}$. It is also used as a 
collective potential in which vibrational states are calculated, 
adding a second contribution to $E_{\text{GCM}}$ of the opposite
(negative) sign.

The total \emph{dynamical correlation energy} is given by
the energy difference between the mean-field ground state and the
projected GCM ground state
\begin{eqnarray}
\label{eq:E:corr}
E_{\text{corr}}
& = & E (q_{\text{mf}}) - E_{k=0}
      \nn \\
& = & E_{J=0} + E_{\text{GCM}} 
.
\end{eqnarray}
Our separation of the dynamical quadrupole correlation energy 
$E_{\text{corr}}$ into a rotational and a vibrational part is 
somewhat arbitrary. An alternative 
choice would be to define the rotational correlation energy 
as the rotational energy of the mean-field ground state 
$E_{\text{rot}}(q_{\text{mf}})$, and take as vibrational energy
the energy gained by the GCM with respect to 
$E_{\text{rot}}(q_{\text{mf}})$. Such a choice would lead to 
smaller rotational but larger vibrational energies, but 
leaves $E_{\text{corr}}$ of course invariant. We prefer the 
separation we have chosen through Eqns.\ (\ref{eq:E:J=0}) and 
(\ref{eq:E:GCM}), that we find easier to interpret.
%
%
\subsection{Two-point topGOA for angular-momentum projection}

The elementary operations of our calculation are the computation
of the overlap $\langle q | \hat{R} (\theta) | q' \rangle$ and the
Hamiltonian  $\langle q | \hat{R} (\theta)\hat{H} | q' \rangle$
matrix elements between two mean-field wave functions
corresponding to different quadrupole moments and to different
orientations in space. For a large-scale calculation as performed
here, it is compulsory to devise an efficient algorithm to reduce
the number of these elementary steps. This can be done at two
places. First, the number of discrete angles $\theta$ necessary to
evaluate the kernels in Eqns.\ (\ref{eq:norm:kernel:exact}) and
(\ref{eq:ham:kernel:exact}) can be reduced using a topological
Gaussian overlap approximation (topGOA) \cite{ha03,Ben04a} for the
$\theta$ dependence of $\langle q | \hat{R} (\theta) | q' \rangle$
and $\langle q | \hat{R} (\theta) \hat{H} | q' \rangle$. Next, the
number of matrix elements to be calculated as a function of $q$
can be reduced by a second GOA, this time for non-diagonal
angular-momentum projected matrix elements between states with
different quadrupole moment.

It has to be stressed that the GOA method as we use it is only a
numerical tool and is quite different from the formal
approximations based on the GOA which are often used in the
literature \cite{Bon90a,Goz85a,Rei87a,Lib99,Yul99a,Pro04a,Fle04a}.
In the framework of this approach, the GOA constitutes the 
first step to
derive a collective Schr{\"o}dinger equation or a microscopic
Bohr-Hamiltonian. By expanding the overlap and energy kernels
around the diagonal matrix elements and assuming a Gaussian shape
in an appropriate set of coordinates, local collective mass
parameters and potentials are derived and used to construct a
collective equation. By contrast, we still solve the projected
HWG equation, Eqn.\ (\ref{eq:HWG}) in our method. Selected matrix 
elements are computed to high precision, allowing us to construct 
a reliable approximation of the full GCM kernels.

A first study of the feasibility of this approach was presented in
Ref.\ \cite{Ben04a}. In the course of the large-scale calculations
reported below, we found that the GOA scheme has to be slightly
modified to ensure convergence of the method in specific nuclei,
mostly light ones and some transitional heavy ones near magic
numbers.

\begin{figure}[t!]
\centerline{\epsfig{figure=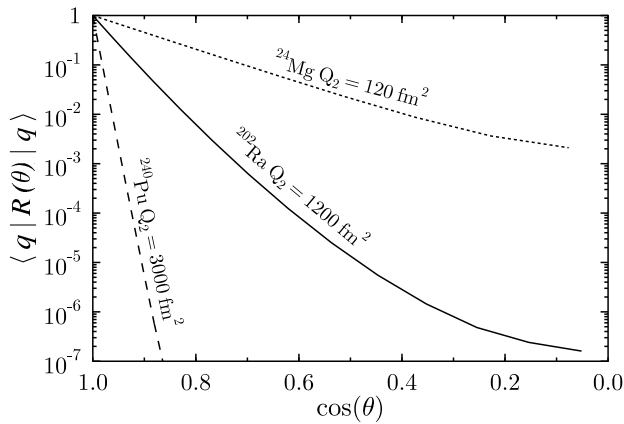}}
\caption{\label{fig:overlap:rot}
Overlap $\langle q | \hat{R} (\theta) | q \rangle$ as a function of
$\cos(\theta)$ for a state close to the projected minimum in
\nuc{24}{Mg}, \nuc{202}{Ra}, and \nuc{240}{Pu}.
}
\end{figure}

With the exception of the cases specified below, a adequate
approximation to the norm and Hamiltonian kernels as a function of
$\theta$ is given by a two-point approximation
\begin{eqnarray}
\langle q | \hat{R} (\theta) | q' \rangle
& = & \langle q | q' \rangle \, e^{-c_2(q,q') \sin^2(\theta)}
      \\
\langle q | \hat{R} (\theta) \hat{H} | q' \rangle
& = & \langle q | q' \rangle \, e^{-c_2(q,q') \sin^2(\theta)} \,
      \nn \\
&   & \times
      \big[   h_0 (q,q')
            - h_2 (q,q') \sin^2(\theta) \big]
,
\end{eqnarray}
where $\langle q | q' \rangle$ is the overlap between unrotated
states and $h_0 (q,q')$ the Hamiltonian kernel between unrotated
states. The width of the Gaussian and the expansion coefficient in
the Hamiltonian kernel are determined from the matrix element
where the left state is rotated by the angle $\theta_2$
\begin{eqnarray}
c_2 (q,q')
& = & \langle q | \hat{R} (\theta_2) | q' \rangle
      \\
h_2 (q,q')
& = & \langle q | \hat{R} (\theta_2) \hat{H} | q' \rangle
.
\end{eqnarray}
A thorough discussion of this particular choice for the GOA and
examples for the quality of this approximation can be found in
Ref.\ \cite{Ben04a}.

The angle $\theta_2$ has to be chosen large enough  to be
sensitive to the variation of the overlap but small enough to fit
the integrand in the region where it brings a large contribution.
Fig.\ \ref{fig:overlap:rot} shows typical overlap functions for
different nuclei.  For magic nuclei and small deformations, the
overlaps do not decrease strongly with $\theta$ while for
\nuc{240}{Pu}, at the ground state deformation, it falls by two
orders of magnitude at a rotation angle of 15$^\circ$. Obviously
the appropriate choice of $\theta_2$ depends on the nucleus. We
determine it from the properties of the mean-field solution,
making use of the approximate overlap function derived by Baye and
Heenen in Ref.\ \cite{Bay84a}. It reduces in our case of axial
nuclei to:
\begin{equation}
\label{eq:ovlp:est}
\langle q | \hat{R} (\theta) | q \rangle
\approx \exp \Big\{ - [ 1 - \cos(\theta) ] \,
                     \langle \hat{J}_\perp^2 \rangle
             \Big\}
,
\end{equation}
where $\langle \hat{J}_\perp^2 \rangle$ is the dispersion of the
angular momentum perpendicular to the symmetry axis of the
mean-field state $| q \rangle$. We choose $\theta_2$ to give an
overlap of 1/2 according to Eqn.\ (\ref{eq:ovlp:est}).  However,
if the equation has no solution, we set \mbox{$\cos(\theta_2) =
1/\sqrt{2}$}. The estimate in Eqn.\ (\ref{eq:ovlp:est}) requires
that the left and right states be the same. For matrix elements
between different states -- and therefore different $\langle
\hat{J}_\perp^2 \rangle$ -- we use the $\langle \hat{J}_\perp^2
\rangle$ that is larger.

Matrix elements between oblate and prolate deformations require
special treatment \cite{Ben04a} because their overlaps peak at
\mbox{$\theta = \pi/2$} rather than \mbox{$\theta = 0$}. These matrix
elements are calculated with a three-point approximation described below.

%
%
\subsection{Three-point topGOA for angular-momentum projection}
For about 10 $\%$ of the nuclei included in this study, often when
the overlap varies slowly with the rotation angle,
a two-point topGOA approximation is not accurate enough, and a
higher-order approximation has to be used. We found a three-point
approximation sufficient in these cases. In the three-point
topGOA, the overlap and Hamiltonian kernels are approximated by
\begin{widetext}
\begin{eqnarray}
\langle q | \hat{R} (\theta) | q' \rangle
& = & \langle q | q' \rangle \,
      e^{-c_2 (q,q') \sin^2(\theta) - c_4 (q,q') \sin^4 (\theta)}
      \\
\langle q | \hat{R} (\theta) \hat{H} | q' \rangle
& = & \langle q | q' \rangle \,
      e^{-c_2 (q,q') \sin^2(\theta) - c_4 (q,q') \sin^4 (\theta)} \,
      \big[   h_0 (q,q')
            - h_2 (q,q') \sin^2(\theta)
            - h_4 (q,q') \sin^4(\theta) \big]
.
\end{eqnarray}
\end{widetext}
The additional parameters $c_4,h_4$ are obtained by demanding
exact values for the angles \mbox{$\theta = 0$}, $\theta_2$,
and \mbox{$\theta_3=\pi/2$}.

\begin{figure}[b!]
\centerline{\epsfig{figure=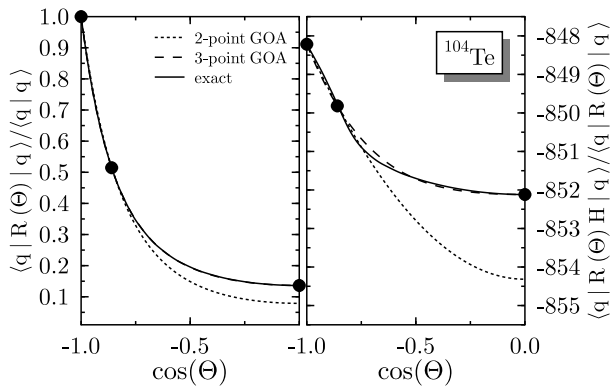}}
\caption{\label{fig:overlap:3pgoa:1}
Comparison between two-point (dotted line) and three-point GOA
(dashed line) and an exact projection (solid line) for the
diagonal overlap (left panel) and  energy (right panel) matrix
elements for \nuc{104}{Te} at a deformation \mbox{$Q_2 = 200$} fm$^2$. A
normalization factor $1/\langle q | \hat{R}(\theta) | q \rangle$ is
included in the energy  that does not enter the calculation of the
Hamiltonian kernel. All matrix elements are projected on particle
number \mbox{$N=Z=52$}. }
\end{figure}

\begin{table}[b!]
\caption{\label{tab:conf:topGOA}
Comparison between different levels of approximation for the \mbox{$J=0$}
norm and energy of the \mbox{$Q_2 = 200$} fm$^2$ mean-field state of
\nuc{104}{Te}.
}
\begin{tabular}{lcc}
\hline \noalign{\smallskip}
method             & $\langle q | \hat{P}^{0}_{00} | q \rangle$
                   & $\langle q | \hat{P}^{0}_{00} \hat{H}| q \rangle/
                      \langle q | \hat{P}^{0}_{00} | q \rangle$ \\
\noalign{\smallskip} \hline \noalign{\smallskip}
two-point   topGOA & 0.0618 & $-850.753$ \\
three-point topGOA & 0.0705 & $-850.487$ \\
full projection    & 0.0706 & $-850.488$ \\
\noalign{\smallskip} \hline
\end{tabular}
\end{table}

In Fig.\ \ref{fig:overlap:3pgoa:1}, the overlap and energy functions
calculated with the two-point and three-point topGOA are compared
with the exact ones. The example
chosen is a case for which the two-point approximation is inadequate,
the diagonal matrix element at \mbox{$Q_2 = 200$} fm$^2$ in the nucleus
\nuc{104}{Te}. The rotation angles used for the topGOA are shown as
dots.  The two-point approximation clearly underestimates the
overlap at large angles, which leads to too small a
projected overlap. The three-point topGOA, on the other hand,
cannot be distinguished from the exact calculation within the
resolution of the plot.
The Hamiltonian matrix element is also underestimated at large
angles, by an even greater amount. This is shown in the
right panel of Fig.\ \ref{fig:overlap:3pgoa:1}.  One also sees
that three-point approximation has only a small difference from
the exact function.

The values obtained for the projection on \mbox{$J=0$} of the
overlap and the energy are given in Table \ref{tab:conf:topGOA}.
As expected, the overlap and the energy obtained with a two-point
topGOA approximation are significantly too small, while the
three-point approximation agrees perfectly with the exact result.

The three-point GOA has been used for nuclei with less than 22
neutrons or protons and for configurations with very small prolate
or oblate deformations. The most critical heavy nuclei differ from
magic numbers by two nucleons or by an $\alpha$ particle. All
other matrix elements were calculated as described in Ref.\
\cite{Ben04a}.

%
%
\subsection{Mixing deformations }
\label{sect:mixing}

The next step in treating correlations by the GCM is to select a set
of deformed configurations, compute the required matrix elements
in Eqn.\ (\ref{eq:HWG}), and diagonalize the corresponding
eigenvalue problem.

Given a set of $n_c$ configurations, the number of overlap or
Hamiltonian computations needed to generate a matrix for Eqn.\
(\ref{eq:HWG}) is $n_c(n_c+1)/2$. As explained in Ref.\
\cite{Ben04a}, the effort can be drastically reduced, to linear
order in $n_c$, by using a topological GOA in the deformation
coordinate.  In the simple case, this requires only calculation of
diagonal and nearest-neighbor off-diagonal matrix elements, i.e.\
\mbox{$2 n_c - 1$} elements per matrix. A subtlety that arises is
that \mbox{$Q_2=0$} is a singular point for the GOA. This does not
cause any difficulty if the nucleus is well deformed either
prolate or oblate, but must be dealt with if the ground state wave
function has significant mixing between prolate and oblate
deformations. We drop the \mbox{$Q_2=0$} configuration which is
nearly redundant with our configuration spaces. We believe that
the correlation energies have a numerical accuracy of about 200
keV with respect to a fully converged GCM.

Some typical configuration sets for heavy nuclei are shown in
Table \ref{tab:conf:space}. The table enumerates the
configurations used for the GCM in those nuclei. The points are
not equidistant, but selected in such a manner that they resolve
the structures in the \mbox{$J=0$} potential energy curve, and
asking that the overlaps between neighboring configurations are
above 0.5 and, if possible, below 0.7. According to Ref.\
\cite{Bon90a}, this range is sufficient to produce final energies
having errors of less than 200 keV. In some cases, however, we
have to add points with larger overlap to ensure that we represent
all structures in the potential landscape. For spherical nuclei
the selection of points requires some search by trial and error. A
set of deformations that can be used for a nucleus with a given
structure, however, works usually also for adjacent ones of the
same type. We usually include a few more points than necessary to
obtain convergence of the GCM ground-state energy.

Numerical stability of the eigenvalue problem is also an issue in these
computations. For a given set of deformations, we always diagonalize
the overlap matrix first, and then remove by trial and error the states
with lowest norm eigenvalues until we obtain a stable solution of the
HWG equation that is not contaminated by spurious states. In some cases
the selection of deformations has to be modified to remove spurious
states.

\begin{table}[t!]
\caption{\label{tab:conf:space}
Configuration spaces for typical heavy nuclei. Mass quadrupole moments
$q$ are given in barns. Oblate and prolate configurations are listed
on separate lines.
}
\begin{tabular}{lcc}
\hline \noalign{\smallskip}
nucleus   &  $n_c$ & $q$ values  \\
\noalign{\smallskip}\hline\noalign{\smallskip}
\nuc{208}{Pb} & 12 & $-20$ $-15$ $-10$ $-7.5$ $-5$ $-2.5$ \\
              &    & +2.5 +5 +7.5 +10 +15 +20\\
\noalign{\smallskip}\hline\noalign{\smallskip}
\nuc{180}{Hg} & 17 & $-24$ $-20$ $-16$ $-14$ $-10$ $-6$ $-4$ \\
              &    & +4 +6 +8 +12 +16 +20 +24 +28 +32 +36 \\
\noalign{\smallskip}\hline\noalign{\smallskip}
\nuc{170}{Hf} & 14 & $-24$ $-20$ $-16$ $-13.75$ $-10$ $-5$ \\
              &    & +5 +10 +15 +19.25 +22 +25 +30 +35\\
\noalign{\smallskip} \hline
\end{tabular}
\end{table}

%
%
\subsection{Assessment of the numerics}

Thanks to the use of the numerical approximations listed in the
previous section, a huge factor
in computing time is gained without significant loss of accuracy.
Projection on angular momentum requires $n_j$  Euler angles (5 to
15) and the GCM mixing in quadrupole moment $n_c$ deformed states
(7 to 25). This gives $n_j \times n_c (n_c + 1) /2\approx 150$-5000
matrix element evaluations altogether.
Our numerical GOA saves a factor of 2 to 3 on $n_j$ as well as
a much larger factor on completing the Hill-Wheeler matrices.
We end up having about
$n_j \times [ n_c + (n_c-1) ] \approx 26$-100 matrix elements
only to calculate exactly. The GOA as we have done it is
designed to describe accurately the correlations in the $0^+$
ground state and most information for spectroscopy is lost. Note
that particle-number projection is still performed exactly. Our
numerical procedure is tuned to achieve a total accuracy better
than 300 keV. This is sufficient for a study of the systematics of
quadrupole correlation energies, which are an order of magnitude
larger.

%
%
\section{Selected examples}
\label{sect:examples}

\begin{figure}[t!]
\centerline{\epsfig{figure=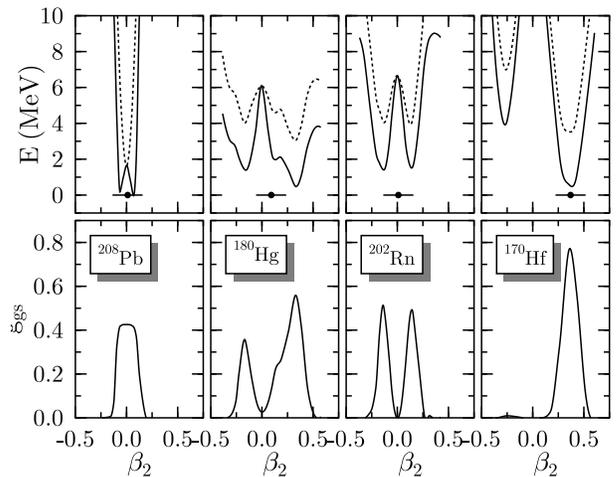}}
\caption{\label{fig:landscape:heavy}
Upper panel: Topology of
unprojected/projected energy landscapes of typical heavy nuclei.
The dotted line denotes the energy after projection on
particle-number only, the solid line the energy after projection
and particle-number and angular-momentum \mbox{$J=0$}. The dot
denotes the energy of the \mbox{$J=0$} projected GCM ground state.
Lower panel: Collective \mbox{$J=0$} ground-state wave function.
All curves and markers are drawn versus the average axial
quadrupole deformation of the mean-field states they are
constructed from.}.
\end{figure}

Figure \ref{fig:landscape:heavy} shows the energy curves (top) and
the collective wave functions (bottom) obtained for cases
representatives of the topographies that one encounters in heavy
nuclei: spherical, soft and  well deformed. The curves are plotted
as a function of a dimensionless axial mass quadrupole deformation
$\beta_2$ defined by:
\begin{equation}
\label{beta2}
\beta_2
= \sqrt{\frac{5}{16 \pi}} \, \frac{4 \pi}{3 R^2 A} \,
  \langle \hat{Q}_{2} \rangle
\end{equation}
with \mbox{$R = 1.2 \, A^{1/3}$}. After angular-momentum projection, we
still use the $\beta_2$ value of the (unprojected) mean-field state
to label the projected \mbox{$J=0$} states, although all \mbox{$J=0$}
states have a zero quadrupole moment in the laboratory frame. The
energy curves projected on \mbox{$J=0$} are also shown in Fig.\
\ref{fig:landscape:heavy}. Finally, a circle in the middle of a bar
indicates the mean deformation of the GCM states, defined as:
\begin{equation}
\label{eq:bar:beta}
\bar\beta_2
= \int \! d\beta_2 \; \beta_2 \; g_{J,k}^2(\beta_2)
.
\end{equation}
The doubly-magic \nuc{208}{Pb} exhibits a very stiff potential
energy curve. Angular momentum projection on \mbox{$J=0$} does not
change that overall behavior, but shifts the minimum of the
potential energy curve to a small, but finite, deformation, a
common feature for all angular-momentum projected energy surfaces
of spherical nuclei \cite{Ben04a,Egi04a}. A spherical mean-field
state is already a \mbox{$J=0$} state, and therefore not at all
affected by projection. Projecting the \mbox{$J=0$} component from
a slightly deformed state, usually with $|\beta_2|$ values below
0.1, often leads to a substantial energy gain, 1.7 MeV in the case
of \nuc{208}{Pb}. Imposing axial symmetry, as done here, the
projection generates two minima which are nearly degenerate and
have similar deformation. The overlaps $\langle J q | J -q
\rangle$ between these minima are very close to one: 0.91 in the
case of \nuc{208}{Pb}. These large overlaps show the limits of
labelling projected states by $\beta_2$: the \mbox{$J=0$} states
obtained by projecting slightly oblate and prolate configurations
are nearly identical. They have the same weight in the GCM ground
state of a spherical nucleus. One of them is, in principle,
redundant, a familiar feature when working in a basis of
non-orthogonal states. The energy gain from the mixing of
different deformations is very small, around 100 keV.

\nuc{180}{Hg} is an example of a transitional nucleus showing
shape coexistence. The mean-field curve presents two  minima,
prolate and oblate, the corresponding wave functions having a very
small overlap, of the order of $10^{-5}$. The energy gain by
angular momentum projection is somewhat larger than in
\nuc{208}{Pb}, but the overall shape of the potential landscape is
not altered by angular-momentum projection. There is also a large
spreading of the collective ground state wave function which, in
particular, mixes prolate and oblate shapes. The energy gain from
the configuration mixing is relatively small, 0.5 MeV only.

\nuc{170}{Hf} is a well-deformed nucleus from the upper end of the
rare-earth region. The mean-field energy curve presents a deep
prolate minimum, the static deformation of the nucleus bringing an
energy gain of 12.2 MeV. Projection of the mean-field energy curve
on \mbox{$J=0$} does not modify the deformation of the minimum but
leads to a gain in binding energy of 2.9 MeV and the GCM mixing of
shapes an extra 0.5 MeV.

The topography of the surface for \nuc{202}{Rn} is intermediate
between \nuc{170}{Hf} and \nuc{180}{Hg}, with two well defined
mean-field oblate and prolate minima of moderate deformations
which are still present after projection. The configuration mixing
gives nearly equal weights to the oblate and prolate deformations,
as can be seen on the collective wave function, the value of
$\bar\beta_2$ being close to zero.

\begin{figure}[t!]
\centerline{\epsfig{figure=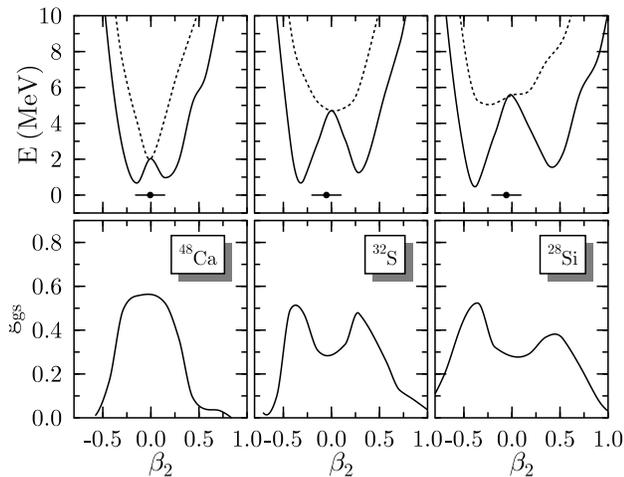}}
\caption{\label{fig:landscape:light}
The same as \protect\ref{fig:landscape:heavy}, but
for light nuclei.
}
\end{figure}

\begin{table}[t!]
\caption{\label{tab:corr:e}
Quadrupolar deformation and correlation energies in
MeV of the nuclei in Figs.\ \ref{fig:landscape:heavy}
and \ref{fig:landscape:light} (see text).
}
\begin{tabular}{lcccc}
\hline \noalign{\smallskip}
nucleus & $E_{\text{def}}$
        & $E_{J=0}$
        & $E_{\text{GCM}}$
        & $E_{\text{corr}}$ \\
\noalign{\smallskip}\hline\noalign{\smallskip}
\nuc{208}{Pb} &  0.0 & 1.7 & 0.0 & 1.7 \\
\nuc{180}{Hg} &  3.0 & 2.6 & 0.5 & 3.1 \\
\nuc{170}{Hf} & 12.2 & 2.9 & 0.5 & 3.4 \\
\nuc{202}{Rn} &  2.6 & 2.7 & 1.4 & 4.0 \\
\noalign{\smallskip}\hline\noalign{\smallskip}
\nuc{48}{Ca}  &  0.0 & 1.4 & 0.7 & 2.0 \\
\nuc{32}{S}   &  0.0 & 3.8 & 0.9 & 4.7 \\
\nuc{28}{Si}  &  0.7 & 4.2 & 0.6 & 4.9 \\
\noalign{\smallskip} \hline
\end{tabular}
\end{table}
%
%

The situation is different in light nuclei, as can be seen in
Fig.\ \ref{fig:landscape:light}. While magic nuclei like
\nuc{48}{Ca} remain stiff and gain only small amounts of dynamical
correlation energy, the ground state of all light open-shell
nuclei is dominated by  dynamical correlations. As fewer
single-particle states cross the Fermi level when deforming light
nuclei, the likelihood to create significantly different
mean-field configurations and coexisting minima is much lower.
With our choice of mean-field and pairing interactions, there are
even only very few light nuclei with a deformed mean-field ground
state (see below). The energy gain from projection is larger than
the static deformation energy, so all non-doubly-magic nuclei have
very similar potential energy landscapes, in most cases with
nearly degenerated prolate and oblate minima, that are strongly mixed by
the GCM. Table \ref{tab:corr:e} summarizes the energy gain at each
step when going from a spherical mean-field state to the
\mbox{$J=0$} projected GCM state.

Some words of caution are necessary here about the vocabulary that
we use to describe our results. Deformation is a well defined
concept for a mean-field state and it can be quantified either by
an intrinsic quadrupole moment or by the parameter $\beta_2$
defined by Eqn.\ (\ref{beta2}). After projection on angular
momentum, a $0^+$ state has of course a zero quadrupole moment
in the laboratory frame.
One can still relate each projected state to a specific mean-field
configuration, and we use it to characterize the projected state.
However, this relation has some limits. First, as very well
illustrated by the case of \nuc{208}{Pb}, the states obtained by
projection of mean-field states with different deformations may
have very large overlap, in particular when they are nearly
spherical. Also, there is no implication that a weakly deformed
configuration has a rotational band. For \nuc{208}{Pb}, for
example, the angular-momentum projected $2^+$ state would 
correspond to a very different expansion of projected mean-field 
states and have a a very different mean deformation $\bar{\beta}_2$.
It is only for cases like the 
well-deformed nucleus \nuc{170}{Hf} that one can expect very similar
collective wave functions for different values of $J$. In
particular, this is the only case where it makes sense to
associate $\bar\beta_2$ with the $B(E2)$ values of the ground
state band as, e.g., in the collective rotor model.

%
%
\section{Overview of correlation energies}
\label{sect:correlations:overview}

%
%
\subsection{Angular momentum projection energies}

\begin{figure}[t!]
\centerline{\epsfig{figure=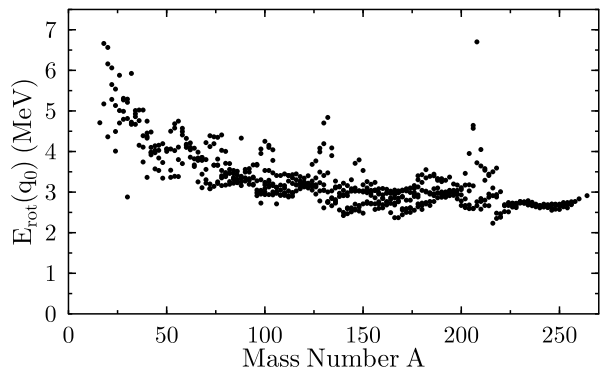}}
\caption{\label{erot_vs_A}
Rotational energy $E_{\text{rot}}(q_0)$ at the minimum of the 
\mbox{$J=0$} projected energy curve.
}
\end{figure}

The angular momentum projection energies $E_{\text{rot}}(q_0)$ are plotted
in Fig.\ \ref{erot_vs_A} as a function of the mass number for the
605 nuclei that we have calculated. One sees that they vary rather
smoothly, decreasing from about 6 MeV in light nuclei to 2.5 MeV
for heavy ones. For a few nuclei around the doubly magic \nuc{132}{Sn} 
and \nuc{208}{Pb}, $E_{\text{rot}}(q_0)$ is much larger and deviate 
from the general trend. In particular, the value of
$E_{\text{rot}}(q_0)$ for \nuc{208}{Pb} is about 7 MeV, but it is largely
compensated by the loss of energy due to deformation. Taking this
loss into account, one obtains the much smaller rotational energy
correction $E_{J=0}$ given in Table \ref{tab:corr:e}.

These projection energies can be compared to values available in
the literature. Egido, Robledo and Rodriguez-Guzman \cite{Egi04a} 
also perform an
exact projection on angular momentum of mean-field wave functions
with an axial symmetry but with another effective interaction, the
Gogny force \cite{RMP} and without projection on particle number.
Girod and coworkers \cite{Lib99} also use the Gogny force and have
developed an approximation scheme for triaxial quadrupole
deformations based on the GCM and leading to a collective
Schr{\"o}dinger equation. Their rotational correction 
is obtained from
the Inglis-Belyaev moment of inertia \cite{Rin80a} but also includes
an \emph{ad hoc} renormalization factor to take into account the
Thouless-Valatin rearrangement contribution. Both Reinhard \etal\
\cite{Rei99a} and Goriely \etal\ \cite{Gor05a} use a
Skyrme effective interaction. The approximation used by Reinhard \etal\
is similar to the one of Girod \etal\ and is derived from
a local GOA approximation of projection. In the work of
Goriely \etal, the rotational correction is determined with a
moment of inertia calculated by a cranking formula, modified either by a
rigid body term or, as in the most recent applications, rescaled at small
deformations to behave in a realistic way.

\begin{table}
\caption{\label{eqJ=0}
Angular momentum projection energies $E_{\text{rot}}$ 
for selected cases compared with other calculations
(see text).
}
\begin{tabular}{lccl}
\hline \noalign{\smallskip}
Nucleus   & $\beta_2$  & This work & Other \\
\noalign{\smallskip} \hline \noalign{\smallskip}
\nuc{32}{Mg}  & $-0.25$ & 4.7 & 6.3 \cite{Rei99a}; 4.0 \cite{Egi04a} \\
\nuc{44}{S}   & $-0.29$ & 4.0 & 6   \cite{Rei99a} \\
\nuc{98}{Zr}  & $-0.11$ & 3.3 & 3.8 \cite{Rei99a} \\
\nuc{164}{Er} & $ 0.36$ & 3.0 & 3   \cite{Egi04a} \\
\nuc{198}{Hg} & $-0.15$ & 3.3 & 3.1 \cite{Lib99} \\
\nuc{240}{Pu} & $ 0.30$ & 2.7 & 3   \cite{Ben04} \\
\noalign{\smallskip} \hline
\end{tabular}
\end{table}

A sample of results obtained with these different methods are
shown in Table \ref{eqJ=0}. The minima of the potential energy
landscapes determined in the five works quoted in Table \ref{eqJ=0} 
might correspond to significantly different deformations since the
effective interactions are not the same. We have therefore
compared the values of the angular momentum projection energies
for the deformation of the minimum that we have obtained here. One
can indeed expect that this energy is not too sensitive to the
details of the interaction and depends mainly on the geometry of
the mean-field wave function which is projected. The values of the
energies obtained by Egido, Robledo and Rodriguez-Guzman with an 
exact projection are rather close to our values. They also obtain very 
similar results for \nuc{208}{Pb}, with a huge energy gain for a small
deformation partly compensated by the loss of energy due to
deformation. The projection energies obtained by Reinhard \etal\
are somewhat larger than ours; the values of Girod \etal\ which
are determined from a very similar method are more similar but
they include an \emph{ad hoc} renormalization factor of 1.32 without
which they would be closer to Reinhard's values than ours. The
inclusion of a rigid body component in the calculations by Goriely
\etal\ also seems important to obtain values close to those of 
an exact projection.

There are unfortunately not many values in the literature
explicitly given for the correlation energies associated with
configuration mixing. Reinhard \etal\ give vibrational energies
for a sample of nuclei which should be an approximation of our
correlation energies $E_{\text{GCM}}$.
Both have indeed the same order of magnitude
and also have the same effect of reducing the increase of
deformation due to the rotational correction. Girod \etal\ treat
these correlations as vibrations in a collective nuclear
potential. This enables them to consider triaxial deformations at
a low cost. However, the variational nature of the GCM is lost in
their approximation and the corrections that they determine are
zero-point energies which increase the energy of the
(approximately) projected ground state.

\begin{figure}[t!]
\centerline{\epsfig{figure=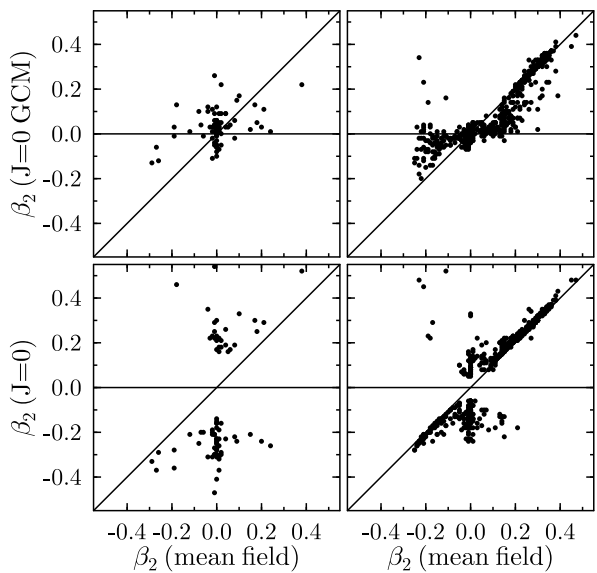}}
\caption{\label{beta_beta}
Mean deformation of the GCM ground state for \mbox{$J=0$} (top) and
deformation of the mean-field configuration corresponding to the
minimum of the \mbox{$J=0$} energy curve (bottom). Both are plotted
as a function of the deformation of the mean-field ground state. 
Left and right panels show light and heavy nuclei, respectively, 
divided at \mbox{$A=60$}.
}
\end{figure}

\begin{figure}[t!]
\centerline{\epsfig{figure=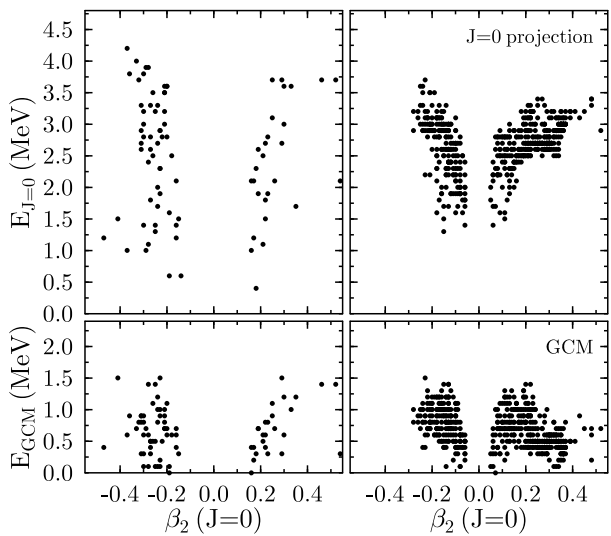}}
\caption{\label{EJ_beta}
Correlation energy $E_{J=0}$ as a function of the deformation 
$\beta_2(q_0)$ of the angular-momentum projected ground state. 
Left and right panels show light and heavy nuclei, respectively, 
divided at \mbox{$A=60$}.
}
\end{figure}

%
%
\subsection{Induced deformations}

Without any exception, the mean-field configuration leading to the
minimum of the projected energy curve is deformed. This is not
surprising; the angular momentum projection of deformed intrinsic
configurations permits the inclusion of small components in the
wave function that might otherwise be treated as perturbative
two-particle two-hole amplitudes \cite{ha03}. In the bottom of
Fig.\ \ref{beta_beta} is plotted the deformation of the minimum of
the \mbox{$J=0$} energy curve as a function of the deformation of
the mean-field ground state. Nuclei are divided into light (left)
and heavy (right) ones. For heavy nuclei, both deformations are
equal as soon as the deformation is of the order of 0.1, with a
few exception corresponding to nuclei with a very soft surface or
a deformed secondary minimum at low energy. For nuclei with masses
lower than 60, both deformations are much more different, with the
general tendency that projection increases the deformation. In
the top panel of the same figure we plot the dependence of the 
mean deformation of the GCM state, as given by Eqn.\ (\ref{eq:bar:beta}),
on the deformation of the mean-field ground state. 
Both quantities are quite close for
heavy nuclei, even when the mean-field ground state is spherical.
For light nuclei, the mean deformation of the projected GCM states 
are also closer to the deformation of the mean-field ground state
than the deformation of the \mbox{$J=0$} projected ground state,
with, however, still large differences.

%
%
\subsection{Systematics of $E_{J=0}$ and $E_{\text{GCM}}$}

Systematics of the correlation energies $E_{J=0}$ and
$E_{\text{GCM}}$ are shown in Fig.\ \ref{EJ_beta} 
as a function of the deformation of the minimum of the
angular-momentum projected energy curve.

Oblate and prolate configurations lead to correlation energies
$E_{J=0}$ of the same magnitude, with a large spreading as a
function of $\beta_2$, smaller for heavy nuclei than for light
ones.  For deformations larger than \mbox{$\beta_2=0.2$}, these
energies vary for  heavy nuclei between 2.5 and 3.5 MeV and for
light ones between 1.0 and 4.2 MeV.

The correlation energy associated with configuration mixing,
$E_{\text{GCM}}$, is plotted in the lower panels of Fig.\ \ref{EJ_beta}.
It is smaller than $E_{J=0}$ with a similar behavior for heavy and 
light nuclei. Although $E_{\text{GCM}}$ could be close to zero for 
some nuclei, it can be as large as 1.5 MeV for others. There is no
clear dependence of $E_{\text{GCM}}$ on the magnitude or on the 
sign of $\beta_2$. Nuclei with small deformation $\beta_2$  of the
mean-field ground state may have correlation energies $E_{\text{GCM}}$ 
as large as very deformed nuclei. It therefore does not seem 
possible to ascribe a dependence of $E_{\text{GCM}}$ on the static
$\beta_2$ value. In collective models, the correlation energy
comes from fluctuations in $\beta_2$. These can only be calculated
by determining the curvature of the energy surface and the
inertial parameter associated with that coordinate. 

%
%
\section{Mass table systematics}
\label{sec:mass:table:syst}

%
%
\subsection{SCMF energies}

\begin{figure}[t!]
\centerline{\epsfig{figure=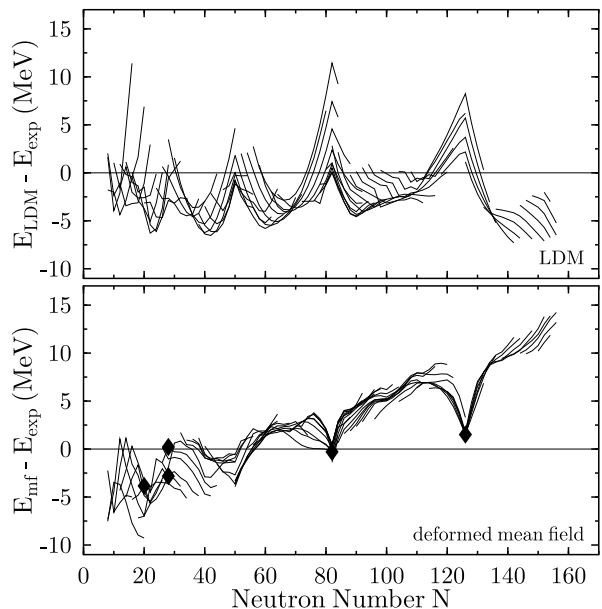}}
\caption{\label{fig:edev:mf}
Deviations between theoretical and experimental energies as
a function of neutron number. The solid lines connect nuclei in
isotopic chains. The liquid drop model, shown in the
top panel, visibly under-binds magic nuclei near \mbox{$N=50$}, 82 and 126.
The bottom panel shows the SCMF for the SLy4 interaction, allowing axial
deformations and including particle number projection. Markers
denote the nuclei used in the fit of the SLy4 interaction.
The numerical precision of the calculations about 1 MeV in heavy
nuclei, with the error mostly proportional to $A$.
}
\end{figure}

Before discussing the binding energy systematics we briefly describe the
systematics of the residuals of binding energies at the SCMF level.
Experimental masses are taken from Ref.\ \cite{Aud03a}. On the lower panel
of Figure \ref{fig:edev:mf} are shown the difference between SCMF energies
and experiment, using the SLy4 functional and pairing defined in
Section \ref{subsect:calc:mf}. The theoretical energies used for
the figure include particle-number projection as well. Positive
deviations denote under-bound nuclei. The plot of the
corresponding residuals for the liquid drop model is shown on the
upper panel. The residuals are obviously correlated in both approaches. 
One clearly sees the magic number effects, with the liquid-drop 
model under-binding magic nuclei, but the SCMF under-binding the
nuclei in between. The residuals for the SCMF might appear large,
but 5 MeV overbinding in light nuclei corresponds to a 3 $\%$ error
on total energies, and 13 MeV under-binding out of nearly 2 GeV
binding energy of a superheavy nucleus is an error of only 0.5 $\%$.
Still, applications of mass formulae to unknown nuclei require a much
better precision.

The usual measure of the quality of a mass model is the rms residual 
energy, defined as
\begin{equation}
\label{eq:sigmarms}
\sigma_{\text{rms}}
= \sqrt{ \frac{1}{N} \sum_{j=1}^{N}
         \big( E_j^{\text{exp}} - E_j^{\text{cal}} \big)^2
       }
.
\end{equation}
The rms deviation $\sigma_{\text{rms}}$ on masses for nuclei
calculated in this study is 5.33 MeV, a value much larger than
what can be achieved by recent HFB mass fits. Similar results have
been obtained for SLy4 and other Skyrme interactions by Stoitsov \etal\
\cite{Sto03a} using a slightly different treatment of pairing
correlations. There are two distinct trends in the deviations:

\begin{itemize}
\item
There is a global trend with $N$, which tilts the median of the
deviations. This overall wrong trend can be removed by a slight
change in the parameters of SLy4, see Ref.\ \cite{Ber05a} and 
Section \ref{sect:global} below. It is
probably an artifact of the fit protocol of the standard Skyrme
interactions, which are adjusted solely to nuclear matter
properties and to the binding energies and radii of a few magic
nuclei. Such a global trend is not present in HFB mass fits
\cite{Ton00a,Gor01a,Sam02a,Gor02a,Sam03a,Gor03b,Gor03a,Sam04a,Gor05a}
done on all known nuclear masses. It is also absent in the recent
relativistic SCMF parameterization DD-ME2 by Lalazissis \etal\
\cite{Lal05a}.
\item
There are several local deviations. Some of them  are obviously
correlated to the spherical magic numbers: the closed-shell nuclei
are over bound relative to the surrounding open-shell nuclei, which gives
rise to characteristic ``arches'' between the shell closures. The
same fluctuations appear in the relativistic SCMF of Ref.\ \cite{Lal05a}.
We will investigate if these local deviations are,
totally or partially, related to dynamical quadrupole correlations
beyond the mean field.
\end{itemize}

The diamonds in Fig.\ \ref{fig:edev:mf} mark the five double-magic
nuclei (\nuc{40}{Ca}, \nuc{48}{Ca}, \nuc{56}{Ni}, \nuc{132}{Sn},
\nuc{208}{Pb}) whose binding energies were included in the fit of
the SLy4 interaction. They all are close to the
\mbox{$E_{\text{mf}}-E_{\text{exp}} = 0$} line, located either on
the bottom of the ravines, or the top of the peaks (\nuc{56}{Ni}) 
in the deviations, which explains why the large deviations seen in
Fig.\ \ref{fig:edev:mf} are not in contradiction with a
least-square fit to the binding energies of a few selected nuclei.

Note that SLy4 has been adjusted to magic nuclei for which there
are no pairing correlations present at the BCS level of approximation.
With our LN+projection scheme, pairing correlations are present
even in doubly magic
nuclei. It increases in particular the binding energy of the
lightest doubly-magic nuclei in the sample of fit nuclei.
\nuc{40}{Ca} and \nuc{48}{Ca} were already over bound by 2.18 MeV
and 1.88 MeV, respectively, and pairing adds about another MeV for
\nuc{48}{Ca} and about 1.5 MeV for \nuc{40}{Ca}.

In section~\ref{subsect:refits}, we shall discuss the effect of
refit of a refit the parameters of the Skyrme interaction to try
to minimize the rms residual or some other measure of the quality
of the theory. With this linear refit of the SLy4 interaction the
rms residual are decreased to 1.8 MeV, much better than the result
of the liquid drop model, but still far from the accuracy of the
theories with additional phenomenological terms in the energy
functional.

%
%
\subsection{Correlation Energies}

\begin{figure}[t!]
\centerline{\epsfig{figure=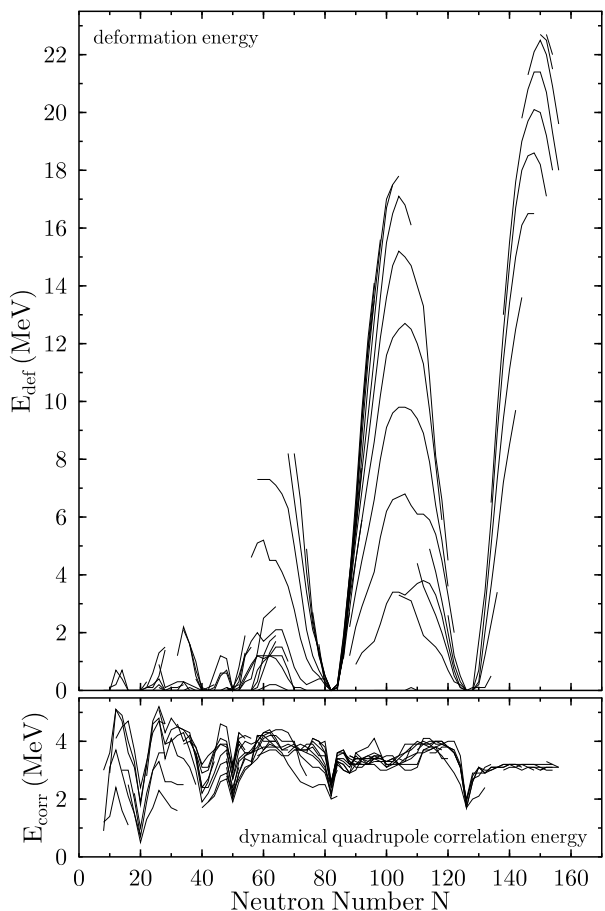}}
\caption{\label{fig:edef_n}
The upper panel shows the static deformation energy has a function
of neutron number $N$. Isotopic chains are connected by
lines. The lower panel gives the correlation
energy including angular momentum projection and mixing deformations.
Note that the panels share the same energy scale.
}
\end{figure}

We have calculated the correlation energies for 605 even-even
nuclei, including the 546 nuclei that have been measured (to a
precision of 200 keV or better). They are available from the
Physical Review archive \cite{epaps} as well as from our own web
site \cite{ev8}.

Figure \ref{fig:edef_n} shows how static and dynamical quadrupole
correlations enter into the total binding energies. In the upper
panel are plotted the static deformation energies. Note that they
include automatically contributions from all multipoles $Q_{\ell 0}$
with even $\ell$.

Both static and dynamic correlation energies are close to zero for
doubly-magic nuclei and increase rapidly away from closed shells
to be maximum mid-shell. In light nuclei, the static correlation
energy never exceeds a few MeV while it grows up to 18 MeV for $A$
between 150 and 180 and  actinides. This energy gain is typical for
non-relativistic interactions, as illustrated by Figure 16
in Ref.\ \cite{RMP}, while it is only around 5 MeV for relativistic
Lagrangians for \nuc{240}{Pu}.
On the other hand, the dynamical correlation energy
is close to 4 MeV for mid-shell nuclei and decreases slightly for
heavy ones. However, static and dynamical correlations behave
differently: the latter are significant as soon as the nucleus is
not a doubly magic one, while the former sets in only in nuclei
with a larger number of protons and neutrons in the open shell.
This has some consequences for mass systematics around closed
shell, as we will discuss it below.

\begin{figure}[t!]
\centerline{\epsfig{figure=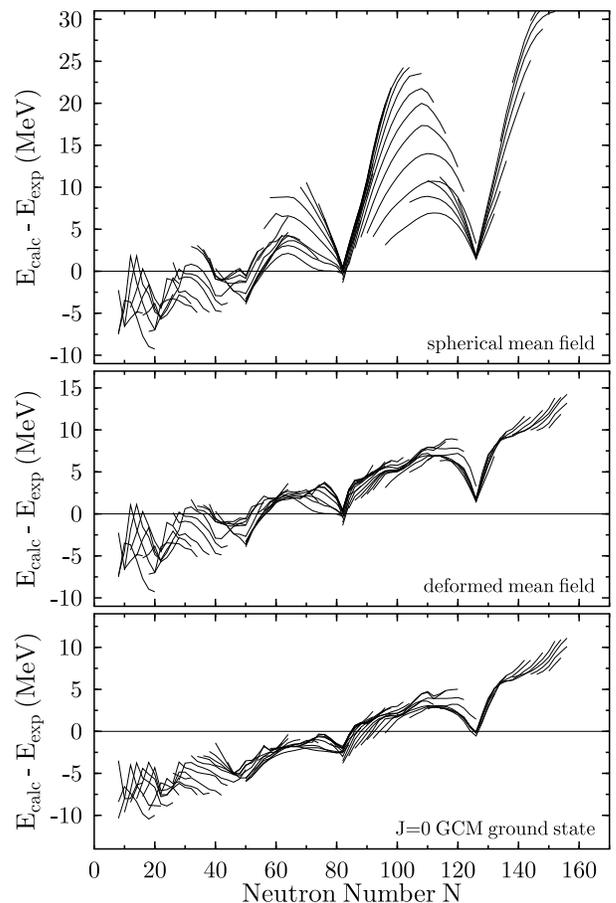}}
\caption{\label{fig:edev_n}
Deviation of spherical mean-field (top panel), deformed mean-field
(middle panel) and \mbox{$J=0$} projected GCM energies from experiment.
Positive residuals denote under-bound nuclei. Note that all panels 
share the same energy scale. Isotopic chains are connected by lines.
}
\end{figure}

The results plotted in Fig.\ \ref{fig:edef_n} are in agreement
with the usual assumption that the mean-field approximation is
better justified in heavy nuclei. For heavy open-shell nuclei with
large symmetry breaking, a large fraction of quadrupole
correlations are static and already included at the mean-field
level. Dynamical correlations dominate the quadrupole energy only
in light systems or around closed shells.

\begin{figure}[t!]
\centerline{\epsfig{figure=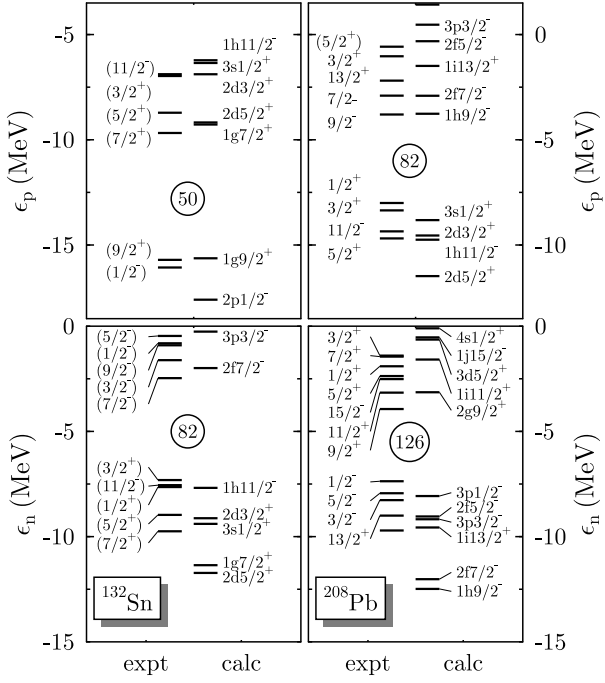}}
\caption{\label{fig:spectra:heavy}
Comparison between calculated (right) and experimental (left)
single-particle spectra of protons (top panel) and neutrons
(bottom panel) for \nuc{132}{Sn} and \nuc{208}{Pb}. See 
Ref.\ \cite{RMP} for the determination of experimental values.
}
\end{figure}

Figure \ref{fig:edev_n} illustrates how the mass residuals
are affected by quadrupole correlations. The top panel shows the
deviation from experiment when spherical symmetry is imposed on
the mean-field. The middle panel, identical to Fig.\ \ref{fig:edev:mf},
includes static correlations by allowing for a deformed mean-field.
In the bottom panel, dynamical correlations from projection on \mbox{$J=0$}
and GCM are included. The difference between the two
upper panels is given by the upper panel in Fig.\ \ref{fig:edef_n},
while the difference between the two lower
panels is given by the lower panel of Fig.\ \ref{fig:edef_n}. As
can be expected from the systematics of deformation energies, to
restrict the mean field to spherical shapes causes huge
fluctuations of the mass residuals for heavy open shell nuclei.
These fluctuations are not removed completely by static deformations, 
but their amplitude and their spread decreases, leaving a plateau for
open-shell nuclei. The curves for all isotopic chains nearly fall
on top of each other. The deviation between  theory and experiment
has now a structure where medium and heavy mid-shell
nuclei fall close to a straight line, while there remain deep
localized ravines around the heavy neutron shell closures
\mbox{$N=50$}, 82 and 126, and more irregular fluctuations in
light systems. Similar results have been obtained for other
effective interactions, see \cite{Sto03a} and references therein.

One can assume that the wrong global trend with $A$ and the deep
ravines around shell closure are correlated to the procedure used to
adjust effective interactions like SLy4. \nuc{208}{Pb} is the only
heavy nucleus included in the fit. A slight error on the volume 
energy coefficient of this interaction leads to an
underestimation of more than 10 MeV of the masses of heavy nuclei. 
Since \nuc{208}{Pb} is the only heavy nucleus included in the fit 
and since its mass is imposed, the error due to the volume energy 
is compensated by a too strong shell effect in \nuc{208}{Pb}.

The additional binding in magic and near-magic nuclei has of course
its origin in shell structure. Hence, single-particle spectra
might offer a key to the understanding of the relative overbinding
of doubly-magic nuclei. The single-particle energies $\epsilon_k$ 
for \nuc{132}{Sn} and \nuc{208}{Pb} are 
shown in Fig.\ \ref{fig:spectra:heavy}. While the  $\epsilon_k$
are not truly physical quantities, there certainly is an 
approximate correspondence to single-nucleon separation energies and 
the spectra of nuclei that differ from doubly-magic ones by one nucleon. 
In this spirit the experimental single-nucleon separation energies
are given as ``expt" in the graph. For a comparison between 
calculated values and experiment one has to keep in mind that
corrections to the $\epsilon_k$ usually increase the level density
around the Fermi energy \cite{Ber80a,Rut98a}.

The SLy4 interaction gives in general a reasonable account of 
the single-particle levels and their ordering around the magic gaps, 
as do most SMCF functionals \cite{RMP}. There are, however, inaccuracies
in details. For example, the magnitude of the gap at \mbox{$N=126$} 
is strongly overestimated, while the gaps at $N$ and $Z$ equal to 82 
appear to be more realistic. The ordering of the levels below the 
\mbox{$N=82$} gap in \nuc{132}{Sn} is difficult to 
reproduce by mean-field models. SLy4 puts the $1h_{11/2^-}$ 
level above the $2d_{3/2^+}$, while experiment gives the opposite 
ordering. SLy4 shares this deficiency with virtually all successful 
parameterizations of Skyrme as well as Gogny interactions and the 
relativistic mean field Lagrangians, see \cite{RMP}. Its consequences
for quadrupole correlations cannot be easily assessed. Another
salient feature of Fig.\ \ref{fig:spectra:heavy} is that the 
calculated level density of neutrons above the \mbox{$N=82$} 
gap in \nuc{132}{Sn} and the \mbox{$N=126$} gap in \nuc{208}{Pb} 
is much lower than the experimental one, which might be one of
the causes for the under-binding of nuclei above \mbox{$N=82$} 
and \mbox{$N=126$}.

The arches that we obtain are still present if the effective
interaction is adjusted to all known masses and has a better
volume energy coefficient. However, the amplitude of the arches 
is much smaller as can be seen, for example, in Fig.~3 of 
Ref.\ \cite{Sam04a}. Dynamical quadrupole correlations reduce 
the fluctuations by approximately a factor 2, suggesting that 
their integration in a global fit of the effective interaction 
might bring a good agreement with the data.

\begin{figure}[t!]
\centerline{\epsfig{figure=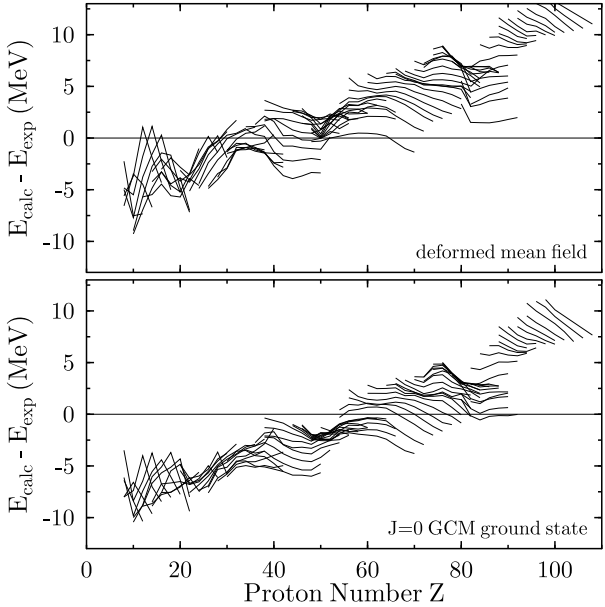}}
\caption{\label{fig:ecorr_z}
Residuals of the (deformed) mean-field energy (top), and
the \mbox{$J=0$} projected GCM energy (bottom) drawn as a function
of proton number. Isotonic chains are connected by solid lines.
}
\end{figure}

Plotting mass residuals for isotopic chains as a function of $N$ is 
the usual way to proceed (see, e.g., Ref.\ \cite{Sto03a}). The plot 
of the same results for isotonic chains as a function of $Z$, however,
leads to a very different perspective on the deviations between theory
and experiment, as can be seen in Fig.\ \ref{fig:ecorr_z}. It demonstrates
that some caution is necessary before drawing conclusions. Although the
lines which connect nuclei with constant $N$ are not perfectly 
horizontal, the fluctuations of the residuals around proton shell 
closures are much smaller at the mean-field level around neutron 
shell closures, and are further reduced when dynamical correlations
are included. The good description of relative energies within a given 
isotonic chain explains why the curves for isotopic chains in 
Fig.\ \ref{fig:edev_n} nearly fall on top of each other. In contrary,
the staggering of the curves for isotonic chains in Fig.\ \ref{fig:ecorr_z}
reflects the drift of the mass residuals along isotopic chains visible
in Fig.\ \ref{fig:edev_n}.

More surprisingly, there is no large missing proton shell effect, hence 
the deep ravines see in Fig.\ \ref{fig:edev_n} are not representative
for shells in general. It is difficult to imagine that close to the stability 
line there are large correlation effects related to neutron shells, 
but not to proton shells. We have seen in Fig.\ \ref{fig:spectra:heavy} 
that single-particle energies of protons are better described 
than those of neutrons. This suggests that the remaining
large fluctuations of the mass residuals around magic neutron numbers
are due to a deficiency of the Skyrme energy functional for 
current parameter sets, and not the manifestation of large missing 
correlations. The arches can still be identified in the mass residuals of
Skyrme interactions fitted to all available masses using approximate
correlation energies, see, e.g., Fig.\ 3 of \cite{Sam04a}.

\begin{figure}[t!]
\centerline{\epsfig{figure=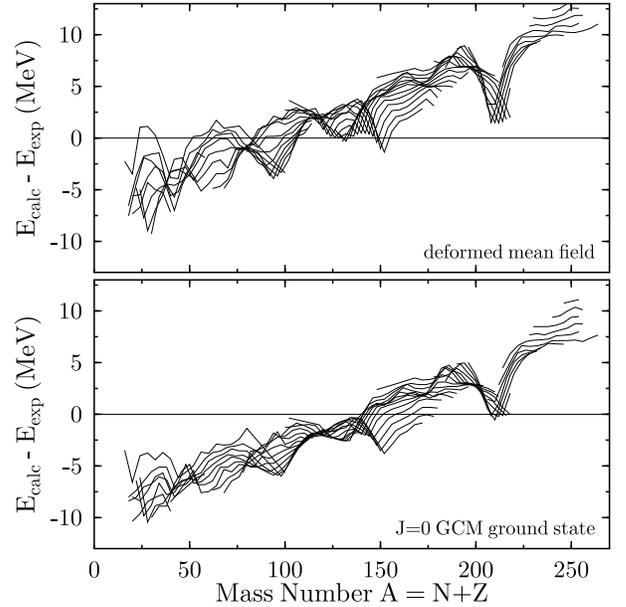}}
\caption{\label{fig:ecorr_a}
Deviation of the (deformed) mean-field energy (top), and
the \mbox{$J=0$} projected GCM energy (bottom) from experiment, 
drawn as a function of proton number. Isotonic chains are 
connected by solid lines.
}
\end{figure}

\begin{figure}[t!]
\centerline{\epsfig{figure=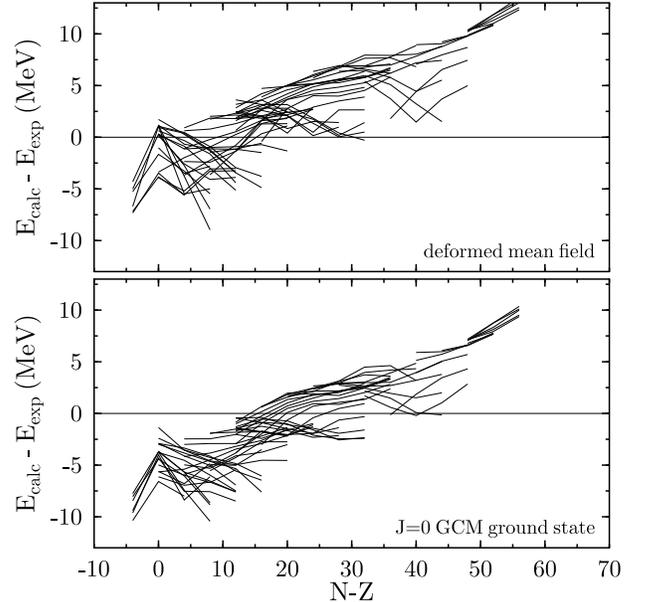}}
\caption{\label{fig:ecorr_i}
Deviation of the (deformed) mean-field energy (top), and
the \mbox{$J=0$} projected GCM energy (bottom) from experiment,
drawn as a function of \mbox{$N-Z$}. Only isobaric chains 
with \mbox{$A=4n$} are drawn, i.e., those containing an 
\mbox{$N=Z$} member. Isobaric chains are connected by solid lines.
}
\end{figure}

The same data also can be drawn versus mass number as in Figures
\ref{fig:ecorr_a}. The lines with constant \mbox{$N-Z$} in Fig.\
\ref{fig:ecorr_a} connect nuclei in $\alpha$-decay chains. The
horizontal lines for the heaviest nuclei indicate that
$Q_{\alpha}$ values are well described, as was first noticed in
Ref.~\cite{Cwi99a}.  Otherwise, the residuals show the same
problems that we found in the plot with respect to neutron number.

Next we show a plot of the isobaric chains as a function of \mbox{$N-Z$},
Fig.\ \ref{fig:ecorr_i}. This clearly shows a strong cusp of the
residuals in light nuclei at \mbox{$N=Z$}, which are
under-bound relatively to the other members of
isobaric chains. It clearly points out that a Wigner energy term is
missing in our model. The amplitude of the fluctuation of the
residuals around \mbox{$N=Z$} suggests a Wigner energy of the
order 5 MeV for the lightest nuclei, and decreasing rapidly with
the mass number. Note that the amplitude of the peaks is modified
when dynamical correlations are included, as the \mbox{$N=Z$} line
contains many mid-shell nuclei that have more correlation energy.

\begin{figure}[t!]
\centerline{\epsfig{figure=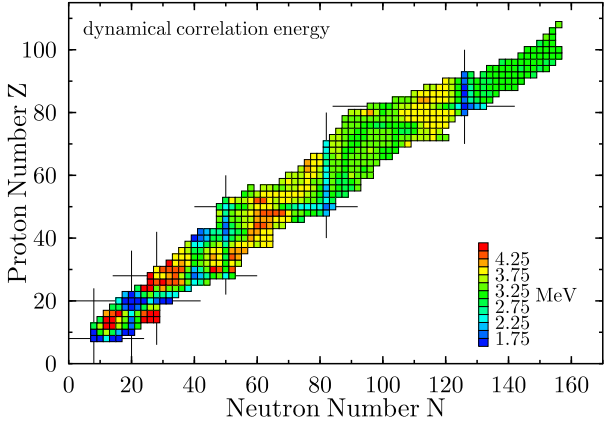}}
\caption{\label{fig:ecorr}
Contour map of the dynamical quadrupole correlation energy in MeV.
}
\end{figure}

Apart from fluctuations correlated to the neutron shell closures,
the trend with $A$ is mainly linear. The refit of SLy4 presented in 
Ref.\ \cite{Ber05a} indeed removes the trend with $A$ with a 0.09 MeV 
increase of the SLy4 volume energy coefficient $a_{\text{vol}}$.
With this increase of $a_{\text{vol}}$, one gains about 21.5 MeV
when going from \nuc{16}{O} to a nucleus with \mbox{$A \approx 250$},
precisely what is needed to correct the slope of the residuals that one
can see in all Figures.

The contour map of the dynamical quadrupole correlation energy is
shown in Figure \ref{fig:ecorr}. It presents structures correlated
to shell effects. The smallest correlation energies are obtained
for magic-nuclei and the largest for transitional nuclei in the
vicinity of shell closures. The maximum of the correlation energy
decreases slightly with $A$. The correlation energy is nearly
constant for rare-earth and actinide nuclei which have all a
static deformation.

The \mbox{$N=28$} shell closure is only clearly visible around
\nuc{48}{Ca}, the dynamical correlation energy increasing very
rapidly when going away from \mbox{$Z=20$}. This result is
consistent with the disappearance of the \mbox{$N=28$} shell
effect below \nuc{48}{Ca}. All the other neutron magic numbers are
predicted to be very stable with only marginal changes of the
correlation energy for each of them.

\begin{widetext}

\begin{figure}[b!]
\centerline{\epsfig{figure=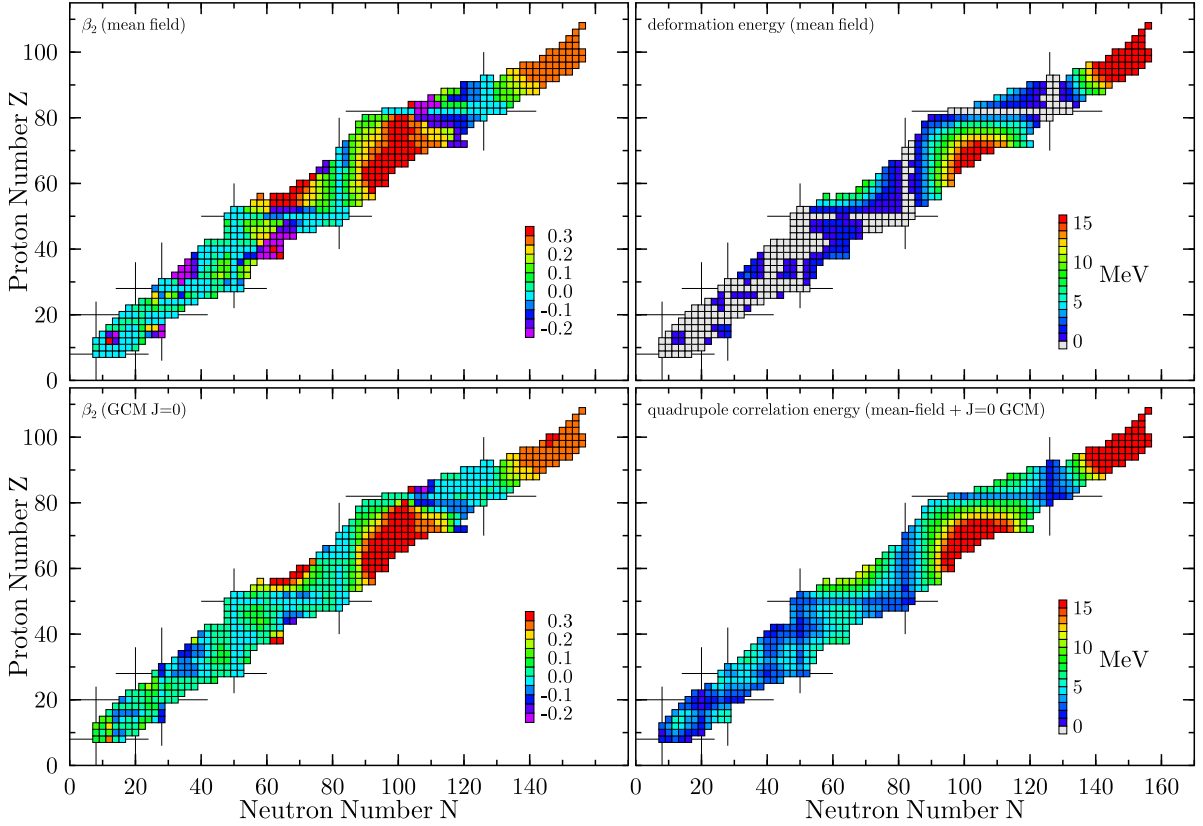}}
\caption{\label{fig:deform}
Left panels: deformation of the mean-field ground state (top) and
average deformation of the \mbox{$J=0$} projected GCM ground state
(see Eqn.\ \ref{eq:bar:beta}). Right panels: static deformation
energy of the mean-field ground state (top) and total
(static+dynamic) correlation energy of the \mbox{$J=0$} projected
GCM ground state. 
}
\end{figure}

\vskip0.5cm

\end{widetext}

The situation is different for proton shells. The correlation
energy is quite small around doubly-magic nuclei, but rises
substantially when going along the shell to mid-neutron-shell
nuclei as one can see for the Ni (\mbox{$Z=28$}), Sn (\mbox{$Z=50$}),
and Pb (\mbox{$Z=82$}) chains. One might suspect that this asymmetry between
neutron and proton shells is an artifact of the too strong neutron
shell effect that has already been noticed. Because of the too
strong neutron shell closures, the potential landscapes are too
stiff, preventing any substantial dynamical correlations. We will
come back to this point when discussing mass differences.

Figure \ref{fig:deform} summarizes the influence of static and
dynamic quadrupole correlations on the ground state wave function
and on its energy. The left panels show the average intrinsic
deformation of the mean-field (top) and of the correlated ground
states (bottom), while on the right panels are plotted the static
deformation energy (top) and the total quadrupole correlation
energy. With SLy4 and our treatment of pairing correlations, most
light nuclei have spherical mean-field ground states (gray squares
in the upper right panel). For nuclei above \mbox{$Z=50$}, there
are three regions clearly visible of well-deformed prolate nuclei
(red squares in the left panels) centered around nuclei that are
mid-shell for protons and neutrons, i.e.\ the rare earths between
\nuc{132}{Sn} and \nuc{208}{Pb}, their cousins with the same $Z$
on the proton-rich side of the \mbox{$N=82$} shell, and the
actinides to the northeast of \nuc{208}{Pb}. The prolate
deformation of rare-earths and actinides is well-established
experimentally. The structure of nuclei with large static
deformation energy is not affected by dynamical correlations. The
situation is different for nuclei at the outer limits of the
deformed regions. There, prolate and oblate, or prolate and
spherical minima coexist and are nearly-degenerate. The GCM ground
state is then a mixing of a large number of configurations, with
an average intrinsic deformation smaller than the mean-field
ground state.

For light nuclei, the mean-field calculations hint at two regions
of well-deformed oblate nuclei with \mbox{$Z \approx 34$} to the
left and right of the \mbox{$N=50$} shell closure. On the
proton-rich side, this is in contradiction with experiment. The Kr
isotopes, for example, are known to have prolate ground states
with coexisting excited oblate structures down to \nuc{74}{Kr};
only \nuc{72}{Kr} has an oblate ground state. In calculations with
SLy4, the oblate minimum is always more bound, which might be
related to a deficiency in the spacing of single-particle states
in the $pf$ shell obtained with this interaction.

Some Sn and Pb isotopes have small ground-state deformation after
projection but before configuration mixing. This is the case when
the mean-field energy surface is soft, a deformed configuration
leading to a small energy gain on the order of 100 keV. After
configuration mixing, however, one obtains a ground-state wave
functions that has zero deformation on the average -- as expected.
We remind the reader again that $\bar \beta_2$ does not have a
physical significance when the deformation is weak.

%
%
\subsection{Mass differences}

In many applications, it is not masses themselves that are
important but differences between masses, as separation energies
or $Q$ values. We have shown that dynamical correlation energies
change abruptly around shell closures and this should have a visible
effect on mass differences. Let us look first to two-nucleon
separation energies
\begin{eqnarray}
S_{2n} (N,Z)
& = & E(N-2,Z) - E(N,Z)
      \nn \\
S_{2p} (N,Z)
& = & E(N,Z-2) - E(N,Z)
.
\end{eqnarray}
They represent first order derivatives of the masses along
isotopic and isotonic chains.

\begin{figure}[t!]
\centerline{\epsfig{figure=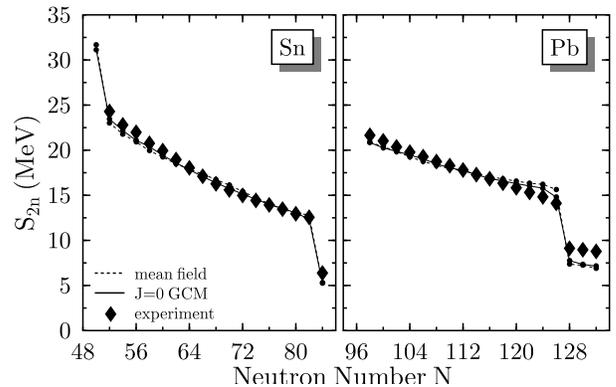}}
\caption{\label{fig:s2n:sn}
Two-neutron separation energy for the Sn and Pb isotopic chains.
}
\end{figure}

Figure \ref{fig:s2n:sn} shows the $S_{2n}$ for the chain of tin
and lead isotopes. All tin isotopes have spherical mean-field
ground states, and the average intrinsic deformation of the \mbox{$J=0$}
GCM states remains close to zero. Dynamical correlations always
bring some gain of energy but which varies slowly for the
open-shell isotopes and energy differences are then marginally
affected. The already good agreement with data at the mean-field
level is slightly improved by correlations, in particular for the
light isotopes and around \mbox{$N=70$}, for which the potential
landscapes are rather soft. The only large change is obtained for
closed-shell nuclei \nuc{100}{Sn} and \nuc{132}{Sn}, for which the
quadrupole correlation energy is smaller than for neighboring
nuclei by about 1 MeV and the ``jump'' in the $S_{2n}$ is reduced
at the shell closure. The $S_{2n}$ in Pb isotopes are also not
much affected by correlations. In particular, the excessive jump
around \mbox{$N=126$} is not sufficiently reduced by correlations.

\begin{figure}[t!]
\centerline{\epsfig{figure=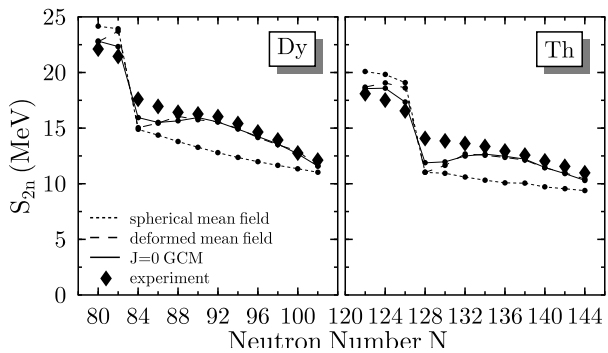}}
\caption{\label{fig:s2n:th}
Two-neutron separation energy for the Dy (\mbox{$Z=66$})
and Th isotopic (\mbox{$Z=90$}) chains.
}
\end{figure}

More complex examples are the chains of Dy and Th isotopes for
which both static and dynamical quadrupole correlations are large.
In both chains, the $S_{2n}$ shown in Fig.\ \ref{fig:s2n:th}
deviate from the experimental values by at least 2 MeV when
spherical symmetry is imposed. Allowing for deformations
significantly improves the agreement with experiment. To obtain
such an effect, the deformation energy has to change by about 2
MeV from one isotope to the next, as it is seen in the upper panel
of Fig.\ \ref{fig:edef_n}. The deformation energy increases
rapidly on both sides of a magic number, but its derivative has a
different sign above and below; therefore deformation decreases
the $S_{2n}$ below a magic number, and increases it above. Going
to $N$ values larger than the magic number, the $S_{2n}$ curve
obtained when deformation is allowed will eventually cross the
spherical curve, when the deformation energy will be decreasing
again with $N$. The dynamical correlations improve the $S_{2n}$
further around the \mbox{$N=82$} (Dy) and \mbox{$N=126$} (Th)
shell closures, in particular below them. Note that the influence
of dynamical correlation energies on separation energies is
necessarily quite localized, as correlations saturate just a few
mass units away from shell closures. The remaining discrepancy
just above the \mbox{$N=126$} shell closure leaves room for
octupole correlations, which are known to be particularly strong
in this mass region \cite{But96a}.

To amplify the change of masses around shell closures even further,
one can study the so-called two-nucleon gaps
\begin{eqnarray}
\label{eq:d2p}
\delta_{2n} (N,Z)
& = & E (N,Z-2) - 2 E (N,Z) + E(N,Z+2)
      \nn \\
\delta_{2p} (N,Z)
& = & E (N-2,Z) - 2 E (N,Z) + E(N+2,Z)
,
      \nn \\
\end{eqnarray}
which are equivalent to second order partial derivatives of masses
as a function of $N$ or $Z$. In a mean-field model,
$\delta_{2p}(Z)$ can be approximated by twice the difference of
the Fermi energies between two nuclei differing by two neutrons or
two protons, provided that all nuclei entering Eqn.\
(\ref{eq:d2p}) have the same structure. For magic nuclei, this
quantity is also approximated by twice the gap in the
single-particle spectrum. For this reason, the two-nucleon gaps
are often used as signatures for magicity.

\begin{figure}[t!]
\centerline{\epsfig{figure=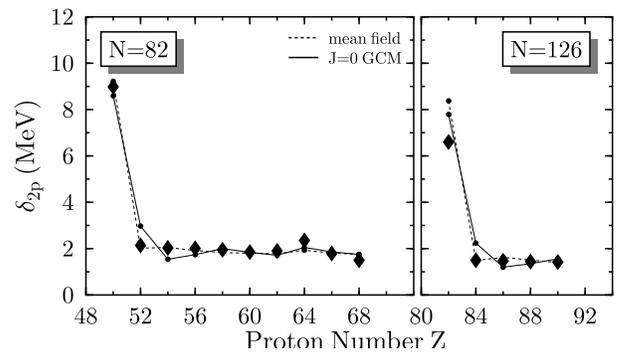}}
\caption{\label{fig:d2p:p}
Two-proton gaps for \mbox{$N=82$} and \mbox{$N=126$} isotonic chains.}
\end{figure}

Figure \ref{fig:d2p:p} shows the two-proton gaps $\delta_{2p}$
along the \mbox{$N=82$} and \mbox{$N=126$} isotonic chains. Except
for the doubly-magic \nuc{132}{Sn} (\mbox{$Z=50$}) and
\nuc{208}{Pb} (\mbox{$Z=82$}), the description of experiment by
mean-field calculations is quite good. Dynamical quadrupole
correlations modify the $\delta_{2p}$ mainly around the proton
shell closure, where the systematics does not necessarily improve.
The $\delta_{2p}$ at the magic number $Z$ decreases, as it does at
$Z+4$, while it increases for $Z+2$. As a result, the mean-field
agreement with experiment at the mean-field level for $Z+2$ and
$Z+4$ is slightly marred by dynamical correlations.

\begin{figure}[t!]
\centerline{\epsfig{figure=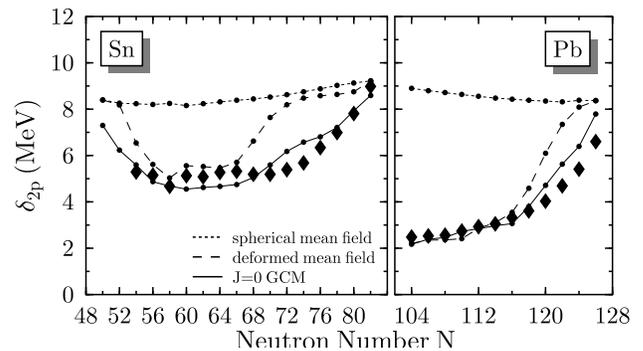}}
\caption{\label{fig:d2p:n}
Two-proton gaps for Pb and Sn isotopic chains. Experimental data
are represented by filled diamonds.}
\end{figure}

The two-particle gaps across a magic number are known to be
difficult to describe by mean-field models. Experimentally,
shell effects are enhanced when both neutrons and protons form
closed shells, a phenomenon called ``mutually enhanced
magicity'' \cite{Zel83a,LMP03} that cannot satisfactorily be described
in a mean-field picture.
It may be seen in Fig.\ \ref{fig:d2p:n}, where the
two-proton gaps for the \mbox{$Z=50$} (Sn) and \mbox{$Z=82$} (Pb)
isotopic chains are plotted as a function of $N$ rather than $Z$.
Now $\delta_{2p}$ represents the magicity of
the proton shell when $N$ is varied. The cases of Sn and Pb have gained
considerable attention, as the experimental data clearly show a large reduction
of the $\delta_{2p}$ when going away from the doubly magic \nuc{132}{Sn}
with 82 neutrons and doubly-magic \nuc{208}{Pb} with 126 neutrons.
The reduction is particularly large for the neutron-deficient Pb
isotopes, which led to some speculations about a possible
quenching of the \mbox{$Z=82$} shell far from stability.

The SCMF gives quite flat predictions for $\delta_{2p}$ in the spherical
approximation, as shown by the dotted line in the figure. This reflects the
independence of the
gap in the single-particle spectrum of the protons on the neutron
number for spherical nuclei \cite{Ben02a}. Allowing for static
deformation leads to a change in the right direction. It is of
course not the Sn and Pb isotopes which gain deformation energy,
but nuclei in the $Z \pm 2$ chains. Adding  dynamical quadrupole
correlations brings the calculated curve very close to the
experimental one. Again, this is due to nuclei in the $Z \pm 2$
chains, which are softer than the magic ones and therefore gain
more dynamic correlation energy. Similar results for the
$\delta_{2p}$ in the Sn chain have been recently obtained with a
microscopic Bohr Hamiltonian based on a different Skyrme
interaction \cite{Fle04a}.

\begin{figure}[t!]
\centerline{\epsfig{figure=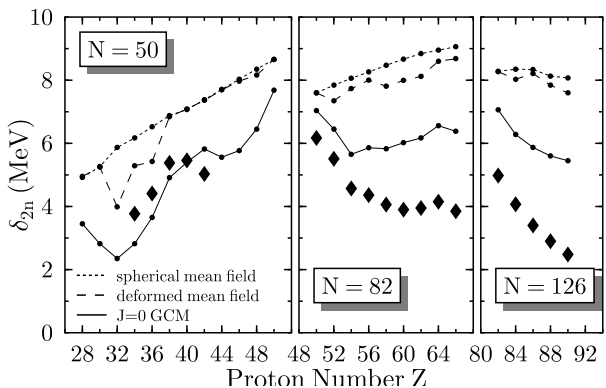}}
\caption{\label{fig:d2n:z}
Two-neutron gaps for \mbox{$N=50$}, \mbox{$N=82$} and
\mbox{$N=126$} isotonic chains.}
\end{figure}

While the correlation energy gives quite a significant qualitative
and quantitative improvement of finite-mass-difference formulae
along the direction of changing proton number, the situation is
less satisfactory along changing neutron numbers. Figure
\ref{fig:d2n:z} shows as an analogue to Fig.\ \ref{fig:d2p:n} for
neutrons the two-neutron shell gap $\delta_{2n}$ across the
\mbox{$N=50$}, \mbox{$N=82$} and \mbox{$N=126$} isotonic chains.
As in the case of proton shells in Sn and Pb, the experimental
values for neutron shell gaps are reduced when going away from the
shell closures. In comparison to the proton case there remain,
however, large deviations from experiment. Values for
$\delta_{2n}$ calculated from spherical mean-field states are flat
only for \mbox{$N=126$}. They vary rapidly for the other two
chains although in the wrong direction with respect to the data
for \mbox{$N=82$}. Allowing for deformation slightly reduces the
$\delta_{2n}$, as some of the nuclei with \mbox{$N \pm 2$} are
deformed. This effect is, however, much weaker than for the proton
shell gaps in the Sn and Pb isotopes. The dynamical correlations
reduce the $\delta_{2n}$ for all nuclei shown by approximately 2
MeV, bringing the theory close to experiment for \mbox{$N=50$}.
This change is, however, not sufficient for the \mbox{$N=82$} and
\mbox{$N=126$} chains. The discrepancy remains the largest for the
\mbox{$N=126$} chain with a much too large neutron shell gap in
\nuc{208}{Pb}.

The failure of the dynamical correlations to describe the
reduction of the $\delta_{2n}$ quantitatively reflects, of course,
the ravines that remain in the lower panel of Fig.\
\ref{fig:edev:mf}. This suggests that the neutron shells in heavy
nuclei are too strong, while proton shells are much better
described. This interpretation is supported by the comparison of
the calculated single-particle spectra of heavy nuclei with
experimental data that we already discussed in Fig.\
\ref{fig:spectra:heavy}.

%
%
\section{Mass-table fits}
\label{sect:global}

In the previous section, we discussed the effects of quadrupole
correlations looking at trends with $N$ and $Z$ and at specific
chains of nuclei. Here, we will take a more global perspective on
the correlation energies, and assess their effect on the table of
nuclei as a whole.

%
%
\subsection{Evolution of errors with the inclusion of correlations}

\begin{table}[t!]
\caption{\label{rms:table:all}
RMS residuals of the binding energy and various binding energy differences
for spherical mean-field states, mean-field ground states, and the
\mbox{$J=0$} projected GCM ground states as obtained with SLy4. 
All energies are in MeV.
}
\begin{tabular}{lcccccc}
\hline \noalign{\smallskip}
Theory & $E$ & $S_{2n}$ & $S_{2p}$
       & $\delta_{2n}$ & $\delta_{2p}$ & $Q_{\alpha}$ \\
\noalign{\smallskip} \hline \noalign{\smallskip}
spherical SCMF     & 11.7 & 1.6 & 1.6 & 1.2 & 1.1 & 2.1 \\
deformed  SCMF     &  5.3 & 1.1 & 1.0 & 1.2 & 1.1 & 1.1 \\
+ \mbox{$J=0$}     &  4.4 & 0.9 & 0.8 & 0.9 & 1.0 & 0.9 \\
+ GCM              &  4.4 & 0.8 & 0.8 & 0.8 & 0.9 & 0.8 \\
\noalign{\smallskip} \hline
\end{tabular}
\end{table}

\begin{figure}[t!]
\centerline{\epsfig{figure=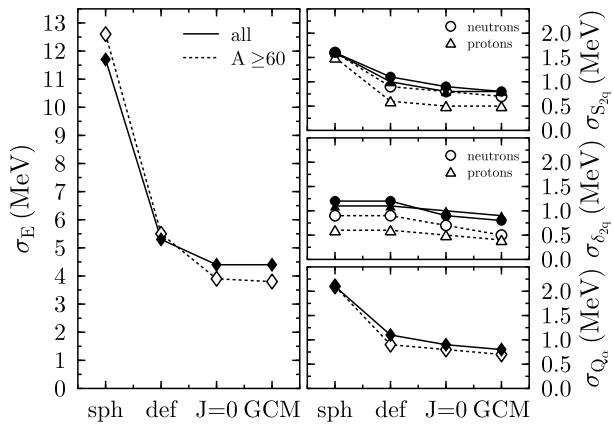}}
\caption{\label{e_rms}
RMS residuals between theory and experiment in different approximations.
Left panel: rms residuals of the masses with the SLy4 interaction
when going from spherical SCMF (sph) to the SCMF ground state
(def), the \mbox{$J=0$} projected minimum (\mbox{$J=0$})
to the \mbox{$J=0$} projected GCM ground state (GCM).
Filled markers denote values for all nuclei in our sample, open
markers heavy nuclei with $N$, \mbox{$Z > 30$} only. The lines are
to guide the eye.
Right panels: corresponding two-nucleon separation energies
$S_{2n}$ and $S_{2p}$ (top), two-nucleon gaps $\delta_{2n}$
and $\delta_{2p}$ (middle) and $Q_\alpha$ values (bottom).
}
\end{figure}

The rms residuals obtained by adding the three components of
quadrupolar correlations to the spherical mean-field values are
given in Table \ref{rms:table:all}. Let us first discuss the
evolution of residuals for binding energies, also shown as the
left panel in Fig.\ \ref{e_rms}. The first line corresponds to
SCMF in the spherical approximation. It has an rms residual of
about 12 MeV. Since the SLy4 interaction is fitted to doubly magic
nuclei, this poor performance in the spherical approximation is to
be expected. Incorporating axial deformations in the SCMF, the rms
residual improves to 5.3 MeV. The next line shows the results
obtained by adding $E_{J=0}$ to the mean-field energies. The
angular momentum projection gives a 20 $\%$ improvement in the rms
residual. This is not surprising; having only fit magic nuclei, a
correlation effect that is stronger for mid-shell nuclei has a good
chance to improve the overall agreement.  The last line shows the
effect of incorporating the full correlation energy. It does not
improve the residuals as compared to the inclusion of $E_{J=0}$
only.

This last result may seem disappointing: the "best" calculation
does not give better residuals for binding energies than
calculations which do not include the correlations due to
configuration mixing. The most obvious factor accountable for this
failure is that the effective interaction that we use has been
adjusted at the mean-field level. Correlation energies always
increase the binding energies. Since with SLy4 light nuclei are
already predicted over-bound by the deformed mean-field
ground state, correlations only worsen the situation, as can be
seen in Fig~\ref{fig:edev_n}. Therefore the correlations cannot 
improve the binding energy residuals.

To check this conclusion, we also give the rms deviations for several
energy differences of interest in Table \ref{rms:table:all}, also shown
in Fig.\ \ref{e_rms}. The spherical mean-field values for $S_{2n}$,
$S_{2p}$ and $Q_{\alpha}$ are substantially improved by static
quadrupole correlations, while the two-nucleon gaps $\delta_{2n}$ and
$\delta_{2p}$ are nearly unaffected. The deviations at the
deformed mean-field level are slightly larger than 1.0 MeV for the
five energy differences. The next two lines of the Table show the
effect of including $E_{J=0}$ and $E_{\text{GCM}}$. In all cases,
one sees a significant improvement ranging from 15 $\%$ to 30 $\%$.
These results are encouraging to demonstrate a role for
correlation effects, but to make a firm conclusion on the need for
the correlation energies one should refit the parameters of the
SCMF and show that the improvement remains when the parameters are
separately optimized with and without the correlations.  We will
come to this in a later section.

For the two-nucleon gaps $\delta_{2n}$ and
$\delta_{2p}$ the static deformation brings no
measurable improvement of the rms residuals, while the dynamical
correlations do. This reflects that the two-nucleon gaps are a
filter for discontinuities in the systematics of masses. The
static deformation energy does not exhibit any noticeable
discontinuities, it only moderately smooths the discontinuity
from the spherical mean field. Therefore it has no visible effect
on the rms residuals of the $\delta_{2n}$ or $\delta_{2p}$.
In contrast, the dynamical correlation energy obviously has a
kink at magic numbers, as illustrated by Fig.\ \ref{fig:edef_n}.
As the most prominent discontinuities of the dynamical
correlation energy and the mean-field coincide, there is
a visible effect of the dynamical quadrupole correlations
on the $\delta_{2n}$ and $\delta_{2p}$.

While the fluctuations of the rms residuals in heavy nuclei are mainly
correlated to magic numbers, they appear to be much more random in
light nuclei, in part due to the additionally missing Wigner energy,
c.f.\ Fig.\ \ref{fig:ecorr_i}. The missing contribution to
the energy appears to reach as far as ten mass units from the \mbox{$N=Z$}
line. As our model cannot describe the effect that leads to the Wigner
energy, we cannot expect to obtain a satisfactory description of light
nuclei. The peak in the energy residuals from the missing Wigner energy
causes a slope and a discontinuity in the energy residuals, therefore
it affects particularly separation energies and two-nucleon gaps.
Indeed, when removing light nuclei with $N$, \mbox{$Z < 30$} from the
calculation of the rms residuals, the overall description of mass differences
appears to be much better, see Table \ref{rms:table:heavy}.

\begin{table}[t!]
\caption{\label{rms:table:heavy}
The same as table \ref{rms:table:all}, but for heavy nuclei with $N$,
\mbox{$Z > 30$} only.
}
\begin{tabular}{lcccccc}
\hline \noalign{\smallskip}
Theory & $E$ & $S_{2n}$ & $S_{2p}$
       & $\delta_{2n}$ & $\delta_{2p}$ & $Q_{\alpha}$ \\
\noalign{\smallskip} \hline \noalign{\smallskip}
spherical SCMF     & 12.6 & 1.6 & 1.5 & 0.9 & 0.6 & 2.1 \\
deformed  SCMF     &  5.5 & 0.9 & 0.6 & 0.9 & 0.6 & 0.9 \\
+ \mbox{$J=0$}     &  3.9 & 0.8 & 0.5 & 0.7 & 0.5 & 0.8 \\
+ GCM              &  3.8 & 0.7 & 0.5 & 0.5 & 0.4 & 0.7 \\
\noalign{\smallskip} \hline
\end{tabular}
\end{table}

%
%
\subsection{Refits of the SCMF interaction}
\label{subsect:refits}

In this subsection we will refit the parameters of the Skyrme
interaction to see the quality of the binding energy fits that can
be achieved with SCMF. Table \ref{rms:table:all} showed that the
correlations are helpful for the SLy4 energy functional, but that
parameterization was never optimized to binding energies. Thus, we
should ask whether the theory does better with the computed
correlation energy if the parameters are optimized by refitting
both cases. As a full fit of a Skyrme interaction to all nuclear
masses is too costly to be performed with the correlation energy
included, we will follow here the procedure of Ref.\ \cite{Ber05a}
to readjust the parameters of the SLy4 functional perturbatively.
Our conclusions will have to be tentative because the pairing part
of the functional was not refitted, and because the perturbatively
refit will not catch a better fit that is very different from the
starting point.

The perturbative refit is performed as follows. The SCMF energy is decomposed
into a sum of integrals, each of which is proportional to some linear
combination of the Skyrme parameters. Because of the variational property
of the self-consistent mean-field theory, these integrals and the
residuals are all that is needed to perform a linear refit of the
Skyrme parameters to minimize the root-mean-square residual or
some other measure of the fit. The only point causing difficulty
is the redundancy of the Skyrme parameters. Certain linear combinations
of these parameters are very poorly determined by nuclear masses,
and should not be included in the fit. According to Ref.\ \cite{Ber05a},
only four combinations out of the ten parameters are well fixed by the
binding energies, and a corresponding singular value decomposition of
the fitting matrix is needed. In Ref.\ \cite{Ber05a}, the energies were
computed with the code {\tt ev8}, but, as explained in Sect.\
\ref{subsect:calc:mf}, the code {\tt promesse} used here is more
accurate.

In Table \ref{fit_table}, we show the results of the fits
optimizing the rms residuals of the binding energies, starting
with the SCMF allowing static deformations. The refitting of the
SLy4 Skyrme parameters gives a very large improvement on the
binding energy residuals, reducing the rms of the SCMF by a factor
of three to 1.8 MeV. The \mbox{$J=0$} projection lowers the rms
residual in a refit by 0.13 MeV, while a refit adding as well
$E_{\text{GCM}}$ lowers it by 0.11 MeV. Thus, a better fit can
be obtained with the correlation energies than without them,
justifying the program of going beyond SCMF in this way. The
situation looks better when we examine fits to mass differences,
which are less sensitive to the SCMF. These are shown in Table
\ref{s2n-refit}. At the level of the SCMF, the effect of a refit
is quite small: 4 $\%$ improvement for the separation energies and
4 to 11 $\%$ for the $\delta_2$. This confirms the common
assumption that the differences are quite insensitive to fine
adjustments of the parameterization. For the theories including
correlation energies $E_{J=0}$ or $E_{\text{corr}}$, the refits
improved the numbers very slightly. Thus, the results shown in
Table \ref{rms:table:all} for the effects of the correlation
energies are also valid when the parameters are readjusted in a
perturbative fit. We emphasize that for the mass differences, both
the angular momentum projection and the GCM mixing of
configurations with different deformation give an improvement.

\begin{table}[t!]
\caption{\label{fit_table}
Quality of binding energy fits for various treatments of effects
beyond mean field. The first line show the SCMF with particle
number projection and below that are the results including
successively particle number angular momentum projection, and
mixing of deformations by the GCM. All energies are in MeV. }
\begin{tabular}{lcc}
\hline \noalign{\smallskip}
Theory                &             rms residual  &      C-norm \\
\noalign{\smallskip} \hline \noalign{\smallskip}
deformed SCMF         &              1.83         &       5.40  \\
+ $J$ projection      &              1.70         &       4.96  \\
+ GCM                 &              1.72         &       5.01  \\
\noalign{\smallskip} \hline
\end{tabular}
\end{table}

\begin{table}[t!]
\caption{\label{s2n-refit}
Quality of fits to separation energies and two-nucleon gaps
fits for various treatments of effects beyond mean field, as
in Table \ref{fit_table}.
}
\begin{tabular}{lcccc}
\hline \noalign{\smallskip}
Theory  & $S_{2n}$ & $S_{2p}$
       & $\delta_{2n}$ & $\delta_{2p}$  \\
\noalign{\smallskip} \hline \noalign{\smallskip}
deformed SCMF         & 1.03  & 0.90  & 1.02  &  1.06 \\
+ $J$ projection      & 0.85  & 0.77  & 0.90  &  0.92 \\
+ GCM                 & 0.79  & 0.72  & 0.83  &  0.84 \\
\noalign{\smallskip} \hline
\end{tabular}
\end{table}

An alternative norm for parameter fitting is the so-called
Chebyshev norm, defined as the largest residual in a fit performed
to minimize that quantity (the ``minimax'' fit).  In Ref.\
\cite{Ber05a} it was shown that the Chebyshev norm could be more
sensitive to the problems in the data set, and it also focuses on
the cases most in need of attention. Fits made by refitting four
well-determined vectors of SLy4 are shown in the last column of
Table \ref{fit_table}. One sees that value of the norm is about three
times the rms value, and that the relative changes from one
treatment of correlations to another are very similar with the two
norms. It is of interest to examine the critical nuclei, namely
the nuclei that have the largest residuals. For an fit with four
independent vectors there are five critical nuclei, given in Table
\ref{tab:critical}.

\begin{table}[t!]
\caption{\label{tab:critical}
Critical nuclei in the perturbative minimax fits to binding
energies starting from SLy4, listed by proton-neutron numbers
$(Z,N)$.
}
\begin{tabular}{lcc}
\hline \noalign{\smallskip}
theory & \multicolumn{2}{c}{critical nuclei}\\
& over-bound & under-bound\\
\noalign{\smallskip} \hline \noalign{\smallskip}
deformed SCMF      & (10,18) (82,126) & (38,38) (38,64) (94,152) \\
+ $J$ projection   & same  & (38,38) (38,64) (42,64) \\
+ GCM              &  (12,8) (82,126) & (38,38) (38,64) (74,108)\\
\noalign{\smallskip} \hline
\end{tabular}
\end{table}

In the mean field approximation, there are two over-bound nuclei among
the critical nuclei, the doubly-magic \nuc{208}{Pb} and the light
neutron-rich nucleus \nuc{28}{Ne}. The under-bound nuclei are two nuclei
along the \mbox{$N=Z$} line and a very heavy neutron-rich nucleus.
With the projections and a refit of the parameters, we expect the
overbinding of magic nuclei to be mitigated, but the change is not
large enough to remove \nuc{208}{Pb} as an over-bound critical nucleus.
The under-bound nuclei are not the same, however.
There is still a nucleus along the \mbox{$N=Z$} line but the others are
middle-mass neutron-rich.  The presence of \mbox{$N=Z$} nuclei in all
lines confirms that special binding effects, the so-called Wigner
energy, are present but not treated by our correlation mechanisms.
We note that the phenomenological treatment of correlations in
Ref.\ \cite{Ton00a} made use of a Wigner energy term with several
parameters.

The continued presence of \nuc{208}{Pb} on the table reflects in
part the special role that this nucleus plays in the construction
of SLy4. A linear refit can correct the global parameters of the
interaction but not the single-particle spectra and it will not
decrease the too large neutron gap. On the other hand, one can
also expect that the inclusion of only quadrupole axial
correlations underestimates the correlation energies and that
other dynamical role should play a role \cite{Hee01a}.

%
%

\section{Systematics of charge radii}
\label{sect:radii}

Correlations also may have an appreciable effect on geometrical
observables such as the mean-square (ms) radius of the charge distribution
$r^2_{\text{c}}$. Although deformation is not a meaningful concept
for $0^+$ states in the laboratory frame, the mean weight $\beta_2$ of
the ground state components can be substantially different from the
deformation of the mean-field ground state. One can therefore expect
particularly large changes of ms radii for light nuclei in general,
and for heavy transitional nuclei.

%
%
\subsection{Procedure}

To calculate the ms radius of the charge distribution, one starts
from the ms radius of the point-proton distribution
\begin{equation}
\label{eq:r:ms:c:GCM}
r^2_{\text{p}}
= \frac{1}{Z} \, \langle \Phi | \, \hat{r}^2_p \, | \Phi \rangle
,
\end{equation}
where $| \Phi \rangle$ is either a mean-field state, a projected
mean-field state or a projected GCM state. To calculate the
non-diagonal matrix element of  $\hat{r}^2_p$, we use the same
method based on the topological GOA as for the Hamiltonian kernel.
A comparison with radii calculated with the complete projected
GCM gives us confidence that the quality of our GOA is on the
order of 0.01 fm, and often much better.

The charge ms radius is then obtained by adding a correction for
the finite size of the proton \cite{RMP}
\begin{equation}
\label{eq:r:ms:c}
r^2_{\text{c}}
= r^2_{p} + 0.64 \, \text{fm}^2
.
\end{equation}
This correction plays no role for differences of charge radii, as
for example isotopic shifts. The root-mean-square (rms) radius is
obtained by taking the square root of Eqn.\ (\ref{eq:r:ms:c}).

For nuclei whose mean-field ground state is spherical, dynamical
correlations always increase the ms radius since deformed
configurations contribute to the collective GCM ground state. For
nuclei with a very shallow deformed mean-field minimum, as for
several heavy transitional nuclei, the GCM ground state will be
spread over a wide range of mean-field states, a mechanism which
could lead to a reduction of the radii.

The amount of increase of radii in spherical nuclei depends on the
softness of the projected energy landscapes. For light nuclei, it
is particularly large, the GCM ground state being spread over a
large range of deformations, as illustrated by Fig.\
\ref{fig:landscape:light}. The charge radii of light nuclei might
increase by more than a percent for the lightest ones, an effect
which should not be neglected if their values are included in the
fit of an effective interaction.

%
%
\subsection{Global Trends}

\begin{figure}[t!]
\centerline{\epsfig{figure=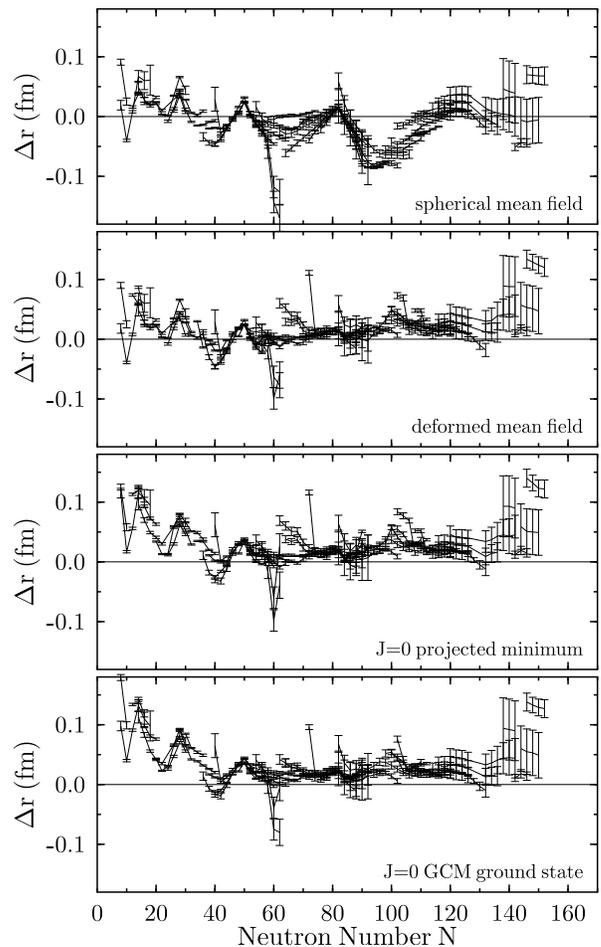}}
\caption{\label{fig:dr:dev:n}
Deviations of the calculated rms charge radii $\sqrt{r^2_{c}}$
from experimental values as a function of neutron number.
We also show the experimental error bars.
Positive values denote overestimated radii. 
Solid lines connect nuclei in isotopic chains.
}
\end{figure}

\begin{figure}[t!]
\centerline{\epsfig{figure=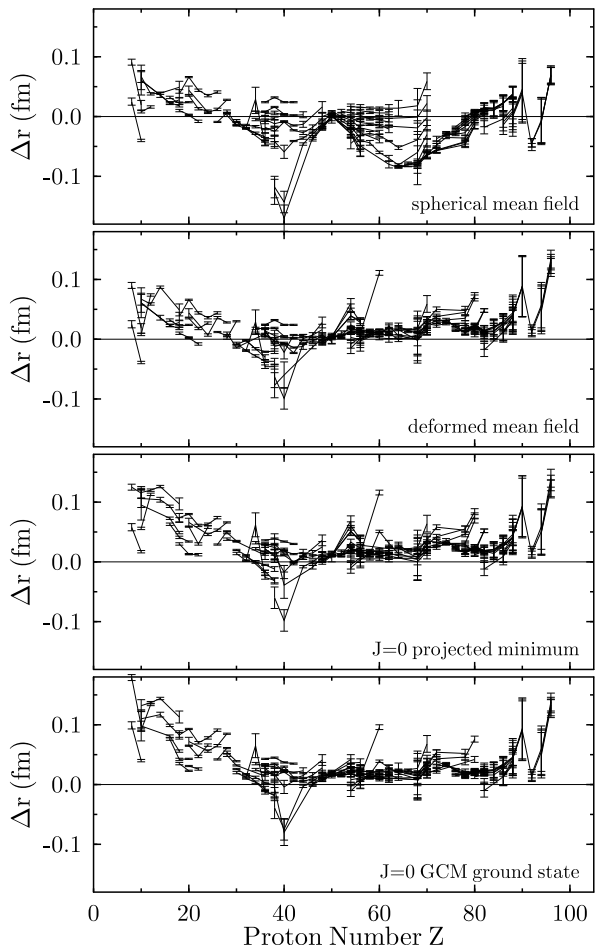}}
\caption{\label{fig:dr:dev:z}
The same as Fig.\ \ref{fig:dr:dev:n}, but plotted for isotonic
chains as a function of proton number.
}
\end{figure}

Figures \ref{fig:dr:dev:n} and \ref{fig:dr:dev:z} show how static
and dynamical correlations influence the deviation of calculated
rms charge radii from experimental data, taken from a recent
compilation by Angeli \cite{Ang04a}. The error bars on rms radii
are often much larger than those on masses. The values for
\mbox{$Z=66$} isotopes with an experimental error bar larger than
0.2 fm have been omitted from the plot. Note that the rms radius
of \nuc{56}{Ni}, that was used for the fit of the SLy4
interaction, is not included in \cite{Ang04a}.

The radii calculated with a mean field restricted to spherical
symmetry underestimate the experimental data for open-shell
nuclei, in particular for the \mbox{$N \approx 66$} region and
rare-earth nuclei \mbox{$N \approx 100$}. This is expected, 
since these nuclei are known to be deformed. Including deformations
improves the agreement with data, as can be seen in the middle
panel of Figs.\ \ref{fig:dr:dev:n} and \ref{fig:dr:dev:z}. In some
mass regions, however, the rms radii are then overestimated, in
particular in the vicinity of the \mbox{$Z=82$} shell closure.
This is again not too surprising, as for transitional nuclei with
soft deformation energy surface, the ground state is poorly
described by a single mean-field state. Dynamical correlations
take into account the spread of the ground state over
deformations, which often reduces the rms radii of many
transitional nuclei. When these correlations are included, the
agreement with data is rather satisfactory for most nuclei in the
region between \nuc{56}{Ni} (\mbox{$N=Z=28$}) and \nuc{208}{Pb}
(\mbox{$N=126$}, \mbox{$Z=82$}), as can be seen in the lower panel
of Figs.\ \ref{fig:dr:dev:n} and \ref{fig:dr:dev:z}. A noteworthy
exception is the neutron-rich Zr region around \nuc{100}{Zr},
where the charge radii are strongly underestimated.

The situation is somewhat different for the lightest and the
heaviest nuclei. The heaviest nuclei in Fig.\ \ref{fig:dr:dev:n}
are known to be well deformed, but for some isotopic chains the
radii are already overestimated by the spherical mean-field
approximation and cannot be improved by the inclusion of
deformations and correlations. Although our results suggest that
SLy4 systematically overestimates the charge radii of heavy
nuclei, a firm statement cannot be made, as, with the exception of
the \mbox{$Z=96$} chain, the experimental error bars are very
large.

For light nuclei, the mean-field ground state is, in most cases,
spherical. The spreading of the GCM ground-state wave functions
over deformation leads then to an increase of the rms radii. As
the radii of \nuc{40}{Ca}, \nuc{48}{Ca}, and \nuc{56}{Ni} are
included in the fit of SLy4 which is done with spherical
mean-field states, their radii are too large when correlations are
included.

\begin{figure}[t!]
\centerline{\epsfig{figure=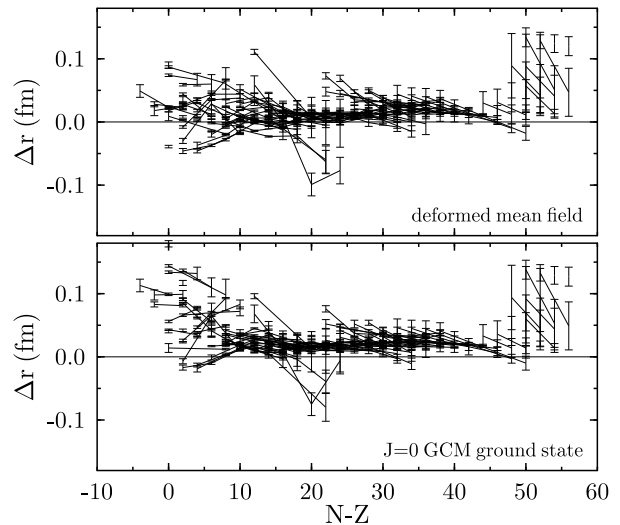}}
\caption{\label{fig:dr:dev:i}
The same as Fig.\ \ref{fig:dr:dev:n}, but plotted for isobaric
chains as a function of \mbox{$N-Z$}.
}
\end{figure}

Residuals for isobaric chains are plotted as a function of isospin
\mbox{$N-Z$} in Fig.\ \ref{fig:dr:dev:i}. For light nuclei, there
is no significant correlation of the residuals with the
\mbox{$N=Z$} line. On the other hand, for heavy nuclei, there is a
clear trend that the theoretical radii do not increase fast enough
with increasing asymmetry \mbox{$N-Z$}.

\begin{table}[t!]
\caption{\label{rms:table:radii:all}
RMS residuals $\sigma_{\text{rms}}$ and average values $\overline{\Delta r}$
of the rms charge radii for all nuclei in our sample (left two columns)
and for heavy nuclei with $N$, \mbox{$Z > 30$} only.
}
\begin{tabular}{lcccc}
\hline \noalign{\smallskip}
       & \multicolumn{2}{c}{all}
       & \multicolumn{2}{c}{$N$, $Z > 30$} \\
Theory & $\sigma_{\text{rms}} (r_{\text{rms}})$ & $\overline{\Delta r}$
       & $\sigma_{\text{rms}} (r_{\text{rms}})$ & $\overline{\Delta r}$ \\
\noalign{\smallskip} \hline \noalign{\smallskip}
spherical          & 0.038 & $-0.012$ & 0.039 & $-0.017$  \\
deformed           & 0.032 &   0.016  & 0.031 &   0.015   \\
+ \mbox{$J=0$}     & 0.041 &   0.027  & 0.034 &   0.022   \\
+ GCM              & 0.043 &   0.030  & 0.033 &   0.023   \\
\noalign{\smallskip} \hline
\end{tabular}
\end{table}

The rms residuals of the radii are given in the first column of
Table \ref{rms:table:radii:all}. There is an overall improvement
when going from spherical to deformed mean-field states. Including
dynamical correlations, however, increases the rms residuals
again. The main reason for this result is that the average value
of the residuals, defined by
\begin{equation}
\label{eq:}
\overline{\Delta r}
= \frac{1}{N}
  \sum_{j=1}^{N} \big( r_{\text{rms}}^{\text{exp}}
                     - r_{\text{rms}}^{\text{cal}} \big)
\end{equation}
and given in the second column of Table \ref{rms:table:radii:all},
increases from 0.016 to 0.031 when adding dynamical correlations
to the SCMF ground state. The overall upward shift when going from
the top panel to the bottom panel in Figs.\ \ref{fig:dr:dev:n} and
\ref{fig:dr:dev:z} reflects the same result. The increase of the
radii is the largest for the lightest nuclei in our sample. As for
masses, the agreement is more satisfactory when removing the
lightest nuclei with $N$, \mbox{$Z \leq 30$} from the sample as
shown in the two right columns of Table \ref{rms:table:radii:all}.

%
%
\subsection{Local trends}

\begin{figure}[t!]
\centerline{\epsfig{figure=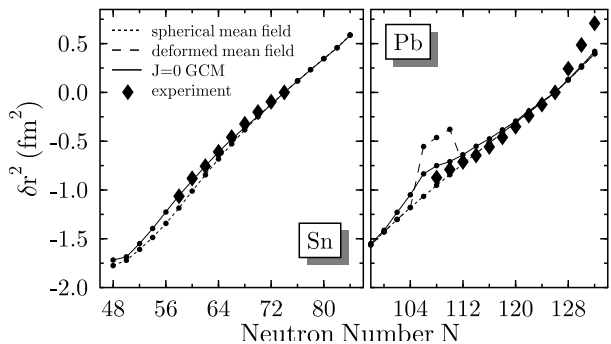}}
\caption{\label{fig:dr2:sn:pb}
Systematics of the isotopic shifts of the mean-square charge radii
along the Sn and Pb isotopic chains. Isotopic shifts of Sn
isotopes are with respect to \nuc{124}{Sn}, in Pb isotopes with
respect to \nuc{208}{Pb}.
}
\end{figure}

As dynamical correlations lead to an overall increase of the rms
radii, any assessment on the role of dynamical correlations for
radii cannot be made on the grounds of the rms residuals. As in
the case of masses, the improvements brought by correlations are
better seen when looking at isotopic shifts which are differences
of radii:
\begin{equation}
\delta r^2_c (Z,N)
= r^2_c (Z,N) - r^2_c (Z,N_0)
.
\end{equation}
In Fig.\ \ref{fig:dr2:sn:pb}, the isotopic shifts for the chains
of Sn and Pb isotopes are compared with experimental data.
Quadrupole correlations play a minor role for Sn isotopes. They
lead to a slight increase for mid-shell nuclei only which improves
the agreement between calculation and experiment. Similar results
have been obtained in a GCM calculation without projection in
Ref~\cite{Benn03a}. The situation is somewhat different for Pb
isotopes. Let us first examine the isotopes below \nuc{208}{Pb}.
For spherical mean fields, the radii vary nearly linearly with
$N$. Allowing for deformation, a few neutron-rich isotopes around
\mbox{$N=108$} have an oblate ground state giving a larger radius
compared to a spherical state. This is seen as a bump in Fig.\
\ref{fig:dr2:sn:pb}. When dynamical correlations are included, the
mean deformation of the correlated ground state decreases and
their radii are intermediate between those of purely spherical and
deformed mean-field calculations. Compared to experiment, there
are some deviations from the linearity of the spherical mean
field, but not as much as predicted by the GCM. Note that the
appearance of a deformed mean-field ground states around
\nuc{186}{Pb} is very sensitive to the strength of the pairing
interaction. Increasing it to $-1250$ MeV fm$^{3}$, as used in
Ref.\ \cite{Ben04b}, pushes the oblate minimum up by a few 100
keV, leading to a spherical mean-field ground state for all Pb
isotopes.
Turning to the isotopes above \nuc{208}{Pb}, one sees a change in
the slope of the isotope shifts in the data that is not reproduced
by theory. Some authors explain the data by using an isospin
dependence of the spin-orbit interaction \cite{Sha95a,Rei95a}
different from that in SLy4 and which appears naturally in
relativistic Lagrangians.

\begin{figure}[t!]
\centerline{\epsfig{figure=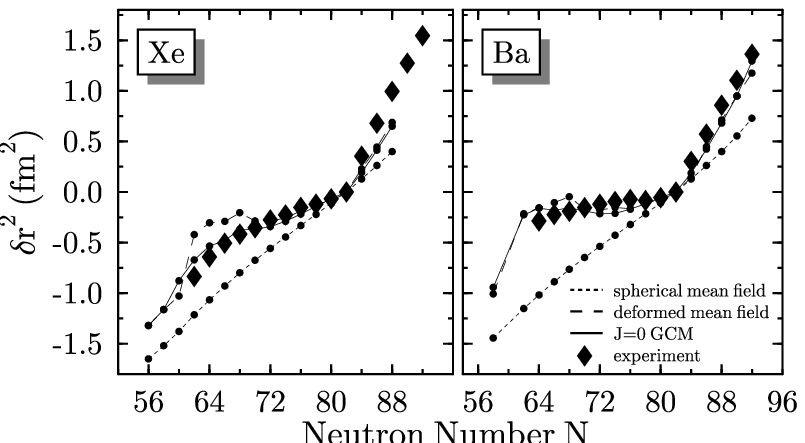}}
\caption{\label{fig:dr2:xe:ba}
Systematics of the isotopic shifts of the mean-square charge radii
along the Xe (\mbox{$Z=54$}) and Ba (\mbox{$Z=56$}) isotopic
chains. Isotopic shifts are with respect to the \mbox{$N=82$}
isotopes, \nuc{136}{Xe} and \nuc{138}{Ba}, respectively.
}
\end{figure}

\begin{figure}[t!]
\centerline{\epsfig{figure=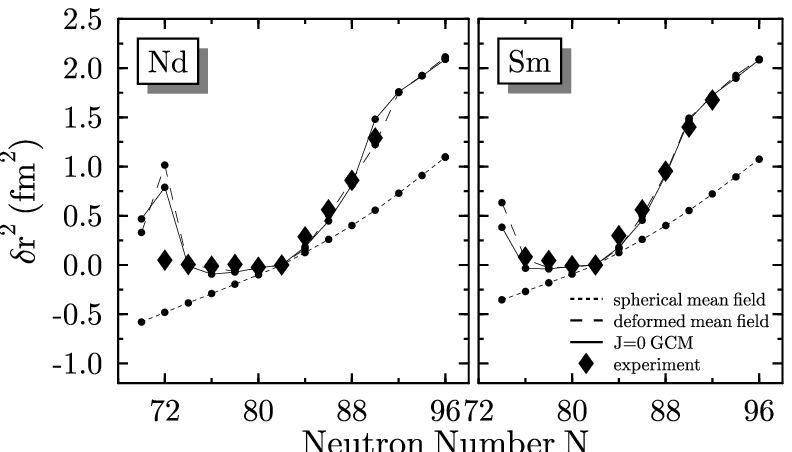}}
\caption{\label{fig:dr2:nd:sm}
Systematics of the isotopic shifts of the mean-square charge radii
along the Nd (\mbox{$Z=60$}) and Sm (\mbox{$Z=62$}) isotopic
chains. Isotopic shifts are with respect to the \mbox{$N=82$}
isotopes, \nuc{142}{Nd} and \nuc{144}{Sm}, respectively.
}
\end{figure}

More drastic changes can be expected for nuclei further away from
shell closures. The isotopic shifts in the Xe (\mbox{$Z=54$}), Ba
(\mbox{$Z=56$}), and Nd (\mbox{$Z=60$}) and Sm (\mbox{$Z=62$})
isotopic chains. are shown in Figures \ref{fig:dr2:xe:ba} and
\ref{fig:dr2:nd:sm} as examples of transitions from near-spherical
to well-deformed ground states along isotopic chains. In all these
cases, spherical mean-field calculations are obviously unable to
describe the trends of the radii. With the exception of the
\mbox{$N=82$} isotopes, all nuclei in Figs.\ \ref{fig:dr2:xe:ba}
and \ref{fig:dr2:nd:sm} are deformed, hence the static quadrupole
correlations in mean-field ground states increase the radii on
both sides of the \mbox{$N=82$} shell, thereby decreasing the
isotope shifts below and increasing them above. The up-bend above
the \mbox{$N=82$} shell seems to be fairly independent on the
proton number and is always well-described. The difference between
the SCMF ground state and the \mbox{$J=0$} projected GCM ground
state is in most cases quite small. Although dynamical
correlations slightly increase them, the isotopic shifts just
above the magic number \mbox{$N=82$} are always slightly
underestimated. The situation is different below the \mbox{$N=82$}
shell. Going from \mbox{$Z=54$} (Xe) to \mbox{$Z=56$} (Ba)
substantially increases the radii for mid-shell nuclei.

\begin{figure}[t!]
\centerline{\epsfig{figure=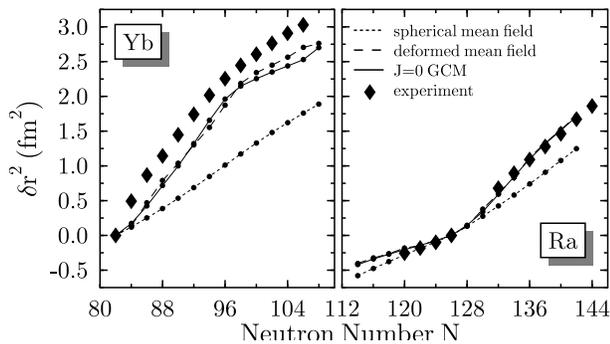}}
\caption{\label{fig:dr2:yb:ra}
Systematics of the isotopic shifts of the mean-square charge radii
along the Yb (\mbox{$Z=70$}) and Ra (\mbox{$Z=88$}) isotopic
chains. Isotopic shifts of Yb isotopes are with respect to
the \mbox{$N=82$} isotope \nuc{152}{Yb}, in Ra with respect to
the \mbox{$N=126$} isotope \nuc{214}{Ra}.
}
\end{figure}

As a final example in Fig.\ \ref {fig:dr2:yb:ra}, we show two
isotopic chains that cross a transition from spherical to strongly
deformed nuclei, Yb (\mbox{$Z=70$}) and Ra (\mbox{$Z=88$}). These
are among the mid-shell chains for which the largest sets of data
are available. A detailed discussion of the isotopic shifts in the
Yb (and also the Pb) chain at the mean-field level can be found in
Ref.\ \cite{Sak01a}. The effect of deformation is clearly visible
in the Yb chain. Static mean-field deformations bring the overall
trend of the radii close to experiment, while dynamical
correlations provide an additional small correction. An
interesting feature is that we find an offset between our
calculations and experiment for most nuclei except the
neutron-magic \nuc{152}{Yb}. The radius does not increase fast
enough when going from \mbox{$N=82$} to \mbox{$N=84$}. This was
already hinted in the case of the Xe, Ba, Nd and Sm isotopic
chains, but it appears in a more pronounced way for Yb. There
could be several sources for this discrepancy. Correlation modes
that we do not consider here should play some role around magic
numbers. Our results on mass systematics have also indicated that
the \mbox{$N=82$} gap is too large. As a consequence, the
potential landscapes are too stiff, preventing the spreading of
the collective ground state.

Compared to the Yb isotopes, the radii of the Ra isotopes vary on
a smaller scale as shown in the right panel of Fig.\
\ref{fig:dr2:yb:ra}. Static deformation increases the radii on
both sides of the \mbox{$N=126$} shell closure, in agreement with
the available data. Adding dynamical correlations leaves the
isotopic shifts practically unchanged.

%
%
\section{Summary, discussion, and outlook}
\label{sect:summary}

Self-consistent mean-field (SCMF) methods provide the only
microscopic nuclear model that can be applied to all nuclei up to
the heaviest ones. In spite of their many successes, the SCMF
models do not deliver nuclear masses with a satisfactory accuracy
if phenomenological corrections are not added. We have studied how
dynamical quadrupole correlations, calculated consistently from
SCMF states, influence masses and charge radii. Using a numerical
approximation to compute the matrix elements required for angular
momentum projection and configuration mixing, it has been possible
to calculate the quadrupole correlation energy for 600 even-even
nuclei. We estimate a numerical uncertainty on correlation
energies of at most 200 keV, which is acceptable for the purpose
of our study.

SCMF masses determined with the Skyrme interaction SLy4  have two
main wrong tendencies: a global drift with mass number, and arches
between shell closures. Similar results have also been obtained
for other Skyrme interactions. The drift with mass number is
related to a slightly too small (approximately by 0.5 $\%$) volume
energy coefficient of the nuclear matter properties, that can be
assumed to be an artifact of the common practice of fitting the
force to a very limited set of nuclei. The mismatch can be easily
removed by a perturbative refit of the particle-hole part of the
effective mean-field interaction to a larger set of nuclear
masses, while leaving the pairing interaction untouched.

The arches are obviously related to shell structure, and cannot be
removed at the mean-field level by a perturbative refit of a given
interaction. Including quadrupolar correlations brings a large
improvement; they have the right qualitative behavior, and also
the right order of magnitude. With the original Skyrme interaction
SLy4, the amplitude of the arches is decreased. The residuals of
the masses remain still large, but in particular mass differences
around magic numbers become rather accurate. The improvements
brought by correlations are smaller when included in a refit, but
clearly are present for angular momentum projection; the overall
improvement of the GCM is small and cannot be determined using a
linear refit.

Surprisingly, the mismatch of masses related to shell effects is
much more pronounced when residuals are plotted for isotopic
chains as a function of the neutron number. In plots made as a
function of the proton number or of the mass asymmetry, the
residuals are much less structured.

The role of dynamical correlations for radii has many similarities
with their role for masses. Correlations lead to an overall
increase of radii that spoils the rms residuals for an interaction
adjusted at the mean-field level. On the other hand, differences
between radii are improved by correlations, the largest effect
being obtained in mass regions where radii vary rapidly from
isotope to the next, in particular for transitional nuclei and
around magic numbers.The examples shown for isotopic shifts
demonstrate that the dynamical quadrupole correlations indeed
contain the right physics to improve the description of the ground
states of transitional nuclei.

The aim of this paper was not to set up a new mass formula,
microscopic and as accurate as the best available theories
including phenomenological terms. This could only be done by
refitting the effective interactions in both the mean-field and
the pairing channels after the inclusion of correlations. Before
to attack this formidable task, our more limited aim here was to
determine the effect of correlations on residuals of binding
energies and to see whether they have the right tendencies to
remove the deficiencies of pure mean-field calculations. Our
results for binding energies are encouraging in this respect.
Energy differences are less sensitive to wrong global trends of
effective interactions and the fact that they are significantly
improved by correlations is a clear sign of the necessity to go
beyond the mean field.

The spreading of the nuclear wave function over a large range of
deformed mean-field states does not only influence binding
energies. It has a large effect on other observables as well. We 
have examined a few representative examples of nuclear radii and
isotopic shifts. Other observables require an extension of the
present study to excited states.

What should be done to go further? We can distinguish several main
roads which should ideally be followed in parallel but which all
require new developments of different difficulties.

\begin{itemize}
\item
From the analysis of mass residuals we conclude that the poor
description of masses around heavy doubly-magic nuclei is due to a
deficiency of the Skyrme energy functional, and is not 
primarily the manifestation of large missing correlations.
This might be a hint that the present Skyrme energy functionals are
not yet flexible enough. To improve the mean field used as a 
starting point, generalizations of the energy functional might 
be necessary. The fit of mean-field interactions should exclude 
\mbox{$N=Z$} nuclei, since the Wigner energy is large and not well known.
Clearly, a fit protocol which takes into account only magic nuclei
is not sufficient.
\item 
the generalization of the present formalism to the study of low-lying 
excited states, in particular to the first $2^+$ state and its decay 
by $E2$ transitions to the ground state.
\item 
the inclusion of additional collective modes. A nice feature of our 
study is that correlations seem to saturate and that the gain of
energy for the ground state is small when introducing more
correlations. However, in some mass regions, they should affect
differently nuclei with different deformation topographies. Three
modes seem the most natural ones looking to some deficiencies
noticed in the present study. The octupole mode is certainly
missing in some heavy nuclei and should also be as important as
the axial quadrupole mode near magic nuclei. The effect of
triaxiality has to be tested in nuclei with coexisting mean-field
prolate and oblate minima which, although already coupled even if
only axial deformations are considered, could be more strongly
linked through the triaxial plane. Finally, pairing has been
treated within the Lipkin-Nogami approach, which is known to have
deficiencies, in particular in the weak pairing regime. To include
the pairing gap as a dynamical variable would certainly be more
satisfactory.
\item 
up to now, we have considered only even nuclei. It is
clear that a more complete theory should also include
odd ones. However, a treatment of odd nuclei at the same level
of quality than even ones will require an extension of our model
to break time-reversal invariance and axial symmetry.
This work is underway but it is quite clear that the restoration
of symmetries for odd nuclei will lead to a considerable increase
of the computing time.
\item 
the discussion of differences of energies and radii demonstrates that
the introduction of quadrupole correlations brings in physics which is 
not included in self-consistent mean-field models and improves the
systematics of ground state observables around shell
closures. However, as we discussed several times in this paper, effective
interactions have been adjusted at the mean-field level and are not
adequate to include correlations. Nuclei whose masses were
included in the determination of the interaction are over-bound
by correlations. A linear refit of the mean-field part of the
interaction permits to correct  wrong tendencies of the
interaction in a simple manner but is
not sufficient for a  quantitative study
when configuration mixing is included. This is not surprising,
since configuration mixing is very sensitive to the relative
position of mean-field energy minima corresponding to different
shapes. One therefore cannot avoid to readjust the effective
interactions in both the mean-field and pairing channels
simultaneously and with the inclusion of correlations.
\end{itemize}

Obviously, these five steps are not completely independent. For
instance, several previous studies have shown that the excitation
energy of the first excited $2^+$ state is overestimated in
spherical or near-spherical nuclei if only an axial quadrupole
collective variable is considered. The variational space should
probably enlarged in these cases by breaking the time-reversal
invariance and introducing a cranking constraint. However,
systematic studies with the present model are still required to
determine which are the critical nuclei requiring an improved
model.

From our study, one can also conclude that the absence of a Wigner
term affects strongly the description of light nuclei, with a
border between light and heavy nuclei around mass 60. The linear
refit of the effective interaction using the C-norm has also put
into evidence critical nuclei. A first way to continue this study
could be to use the C-norm to establish a larger list of critical
nuclei. This will require to extend the data included in the refit
of the interactions in order to lift the redundancy of the
effective interaction parameters.

%
%
\section*{Acknowledgments}
We thank T.\ Duguet and W.\ Nazarewicz for their continuing interest
in this work and many fruitful discussions on the interpretation of
its results. We also thank B. Sabbey for
help with various technical aspects of this work, particularly
on the binding energy fits. PHH thanks the Institute for Nuclear
Theory for its hospitality during the initial stage of this work,
MB thanks the Service de Physique Nucl{\'e}aire Th{\'e}orique at the
Universit{\'e} Libre de Bruxelles for its warm hospitality during
the final stage of of this work.
Financial support was provided by the U.S.\ Dept.\ of Energy under
Grants DE-FG02-00ER41132 (Institute for Nuclear Theory) and
W-31-109-ENG-38 (Argonne National Laboratory),
the US National Science Foundation under grant PHY-0244453,
and the Belgian Science Policy Office under contract PAI P5-07.
The computations were performed at the National Energy Research
Scientific Computing Center, supported by the U.S.\ Dept.\ of
Energy under Contract No.\ DE-AC03-76SF00098.
%
%


\begin{thebibliography}{99}
\bibitem{ev8}
  The SCMF wave functions calculated with the program
  {\tt ev8} are available as well as the correlation energies are
  available at {\tt gene.phys.washington.edu/ev8}.
\end{thebibliography}
\end{document}